\documentclass[10pt,sigconf,letterpaper,nonacm]{acmart}

\usepackage[english]{babel}
\usepackage{blindtext}
\usepackage{subcaption}
\usepackage{booktabs}

\usepackage{graphicx}
\graphicspath{ {./Images/} }
\usepackage{xspace}
\usepackage{xcolor}
\usepackage{balance}
\usepackage{textcomp}
\usepackage{enumitem}
\usepackage{listings}
\usepackage{xspace}
\usepackage{relsize}
\usepackage{environ}

\newcommand{\tool}{\text{UgoVor}\xspace}

\lstdefinelanguage{UgoVor}%
  {
    morekeywords={"windowsize", "rebuffering-levels", "resolution-levels", "720p", "480p",
    "1080p", "4K"}
  , alsoletter=-
  , sensitive=true 
  , morecomment=[l]{\#\ } 
  , columns=fullflexible
  , keepspaces=true
  , fontadjust=true
  , basicstyle=\relscale{0.9}\sffamily 
  , upquote=true
  , showstringspaces=false
  , stringstyle=\color{red!70!brown}
  , escapeinside={(*}{*)}
  , alsoletter={"}
  , literate=
     {<}{$\text{\textless}$}{1} 
     {>}{{$\text{\textgreater}$}}{1}
     {\{\{}{\copen}{1}
     {\}\}}{\cclose}{1}
  }
  [keywords,comments,strings]

\lstnewenvironment{ugovor}[1][]%
  { \minipage{\textwidth}
    \lstset{language=UgoVor,xleftmargin=200pt,#1}
  }
  { \endminipage
  }

\lstMakeShortInline| % allows us to write "code" instead of \lstinline{code}

\NewEnviron{makefitalign}{
  \resizebox{.9\columnwidth}{!}{%
  \begin{minipage}{\columnwidth}
  \begin{align*}
  \BODY
  \end{align*}
  \end{minipage}}
  \medskip
}

\definecolor{Gray}{gray}{0.9}

\renewcommand\footnotetextcopyrightpermission[1]{} % removes footnote with conference info
\setcopyright{none}
\acmDOI{}

\acmISBN{}

\acmConference[]{}
\acmYear{}
\copyrightyear{}

\acmPrice{}

\hypersetup{draft}
\begin{document}
\title{Multilateral Micro-Monitoring for Internet Streaming}

%\titlenote{Produces the permission block, and copyright information}
%\subtitle{Extended Abstract}

\author{Roshan~Shyamsunder, Lukas~Lazerek, Christos~Dimoulas,
Aleksandar~Kuzmanovic}
 \affiliation{%
   \institution{Northwestern University}
 }

\renewcommand{\shortauthors}{R.~Shyamsunder, L.~Lazerek, C.~Dimoulas, A.~Kuzmanovic}

%-------------------------------------------------------------------------------
\begin{abstract}
%-------------------------------------------------------------------------------
Video streaming is dominating the Internet. 
  % with IP video traffic
  %$ contributing to nearly 80\% of global traffic, and with streaming TV
%$taking over traditional cable and satellite options. 
  To compete with the performance of traditional cable and satellite
  options, content providers outsource the content delivery to third-party
  content distribution networks and brokers.  However, no existing
  auditing mechanism offers a multilateral view of a streaming service's
  performance. In other words, no auditing mechanism reflects the mutual
  agreement of content providers, content distributors and end-users alike
  about how well, or not, a service performs. 
  
  In this paper, we present UgoVor, a system for monitoring multilateral
  streaming contracts, that is enforceable descriptions of mutual
  agreements among content providers, content distributors and end-users.
  Our key insight is that real-time  multilateral micro-auditing---capable
  of accounting for every re-buffering event and the resolution of every
  video chunk in a stream---is not only feasible, but an Internet-scalable
  task. To demonstrate this claim we evaluate UgoVor in the context of a
  10-month long experiment, corresponding to over 25 years of streaming
  data, including over 430,000 streaming sessions with clients from over
  1,300 unique ASes. Our measurements confirm that UgoVor can provide an
  accurate distributed performance consensus for Internet streaming, and
can help radically advance existing performance-agnostic pricing models
towards novel and transparent pay-what-you-experience ones.
\end{abstract}
\maketitle

\section{Introduction}
\label{intro}

Video streaming is thriving on the Internet. Globally, video traffic will
be 82\% of the entire Internet traffic by 2022, up from 75\% in
2017~\cite{cisco-wp}.  A large user demand along with advancements in
content access devices, such as tablets, smart TVs, and smartphones, have
all contributed to a dramatic growth of the video streaming applications
in the last decade.  Consequently, significant advances were made, both by
industry and academia, to develop and deploy ever-improving streaming
technologies~\cite{oboe,vantage,salsify,minerva,mpdash,cs2p,pensieve,qflow}.

"Cord-cutting"---a multi-year trend of viewers shifting their "eyeballs
and dollars" from traditional cable and satellite options to streaming
TV--- is becoming a reality~\cite{conviva-report-2018}.  Yet for streaming
TV to cope with the legacy cable and satellite competition, it is
paramount to retain the same level of service.  Specifically users are
shown to be highly impatient with low-quality streaming sessions such as
those that involve rebuffering events, which are known as "silent
engagement killers"~\cite{conviva-report-Q2-2019}.

To achieve the high quality video streaming performance required by
end-users, content providers \emph{outsource} the content delivery to the
content delivery "industry," namely third-party Content Delivery Network
(CDN) providers, \emph{e.g.}, \cite{akamai,alibaba,amazon,google}, and
brokers~\cite{Conviva,citrix}. CDNs typically deploy a large network
infrastructure, thus bringing streaming servers closer to end-users. To
mask ISP performance "glitches," CDNs on the fly,
mid-stream, redirect a user to a different CDN replica, or brokers
redirect a client to a different CDN~\cite{CDNbroker:17}. 

Content providers, CDNs  and other entities regularly collect information 
to audit the performance of streaming services. 
For instance, CDN brokers \cite{Conviva}, largely have
access to end-user streaming health metrics~\cite{conviva-report-Q2-2019}.
Similarly, third-party entities can collect and aggregate client
measurements from many vantage points~\cite{thousandeyes}.  Likewise, CDNs
 deploy client-side plugins to obtain video performance
insights directly from clients~\cite{akamai-media-analytics}.  Content
providers do the same via the video players they distribute to end-users.
However,  \emph{client-only measurements cannot be used for settling payments
between content providers, CDNs and end-users.} Without transparent, consistent,
independent and auditable information from both
end-users and the CDNs, no multilateral \emph{consensus}
about the streaming performance is possible.  
In sum,  \emph{current auditing mechanisms are largely unilateral
and do not use information that is mutually agreed by all
participants in a steaming sessions: the content provider, the CDN
and the end-user}.

The starting point of this paper is that the obstacle for multilateral
stream-auditing is the status quo of coarse-grained agreements 
between content providers, CDNs and end-users. 
Specifically, streaming quality service agreements are in terms of
target aggregate metrics over (long) periods of time, which are difficult
to reconcile when measured from different vantage points.

In response, this paper introduces \emph{micro-auditable
streaming contracts}.  These contracts aim to account for fine-grained
aspects of a video stream's performance such as every rebuffering event or
the resolution of every single video chunk in a stream. Their
micro-auditable  nature renders them \emph{multilateral} and
\emph{enforceable at real-time} by design. Indeed, we construct UgoVor, a
real-time monitoring  system for \emph{micro-auditable streaming contracts}, and demonstrate that
multilateral stream-auditing is a feasible Internet-scale task.

The first challenge for multilateral real-time micro-auditing is
scalability: how can \tool account for every rebuffering event and
monitor the resolution of every video chunk without adding significant
communication overhead between endpoints? In other words, to achieve
scalability, \tool needs to deal with the fundamental information
asymmetry between the clients and servers in streaming systems. Clients
have detailed information about streaming performance metrics; servers, on
the other hand, are "dummy boxes" that simply respond to requests.  To
address the asymmetry, \tool introduces a \emph{virtual buffer} on its
server side monitor, which conservatively approximates  the client buffer state.
Specifically the virtual buffer of the server monitor does \emph{not} ideally
replicate the client's buffer state, yet it includes sufficient
information for the confirmation of video chunk resolution and buffer
health metrics that the client observes. That is, in \tool, the client
monitor,
which has the perfect knowledge, exclusively raises contract violation
challenges.  The server monitor, on the other side, is capable of
deterministically confirming all actual contract violations raised by the
client.  As a result  both clients and servers independently but in tandem
monitor streaming contracts with minimal overhead. 

However, contract monitoring cannot be blindly trusted. Indeed, clients
may try to fabricate contract violations while servers may try to deny
real contract violations. To address this second challenge, UgoVor relies
on a simple tit-for-tat mechanism. In particular, \tool forces client
and server monitors to reach a consensus on the underlying performance measures
over short time scales.  Since the virtual buffer of the server monitor soundly
approximates the state of the client's buffer, consensus is always reached
unless if one of the endpoints is misbehaving. To disincentivize
misbehaving endpoints, when consensus is not reached, UgoVor interrupts
the streaming session. As a result,  both servers and  clients have
limited benefits from being untruthful but a lot to lose. After all, the
CDN does not want to lose a client, and the client does not want to stop
receiving the stream.  In sum, \tool's simple tit-for-tat incentive
mechanism enables truthful reporting without strong identities.

Adoption incentives and ease of deployment are essential for the success
of any Internet-scale system, and strong incentives are indeed present for
streaming contracts. Content providers and advertisers are vitally
interested in auditing how their money is spent, and how are clients
actually being served. CDNs and other streaming providers are incentivized
to support streaming contracts in order to sustain the growing competition
on the streaming delivery market. Finally, end-users likely benefit the
most because their experience is transparently accounted for. As for ease
of deployment, \tool's client monitors can be seamlessly installed via
updates of the content provider's client players. 
In general, \tool is compatible with
existing auditing mechanisms that are already in wide use. All content
providers and CDNs need to do is update their software to implement
UgoVor's tit-for-tat approach to monitoring streaming contracts.

To evaluate UgoVor, we utilize a 10-month long experiment, which
corresponds to over 25 years of streaming data, and includes over 430,000
streaming sessions with clients from over 1,300 unique ASes.  The
collected data demonstrates that streaming quality reduction, manifested
in lower streaming resolution and rebuffering events, is a common case on
the Internet. In particular, we find that nominally High Definition (HD)
streams contain approximately 13\% of non-HD, lower resolution chunks. We
also find that 12.5\% of flows experience at least one rebuffering event,
and that rebuffering events are quite severe, \emph{i.e.}, 95\% of them
are longer than a second.  

Using the experimental data, we show that UgoVor accurately verifies 
streaming resolution and rebuffering events from the experimental data.  
Most importantly, we demonstrate that UgoVor
scales. In particular, there exists \emph{no} difference in the service
quality experienced by users when UgoVor is deployed vs. when it is not.
The most significant increase in CPU utilization between video servers
with and without UgoVor is 4.05\%, while the network bandwidth increase
with UgoVor is 0.51\% in the worst case scenario.  We achieve this via a
careful design and implementation at endpoints -- 
by decoupling traffic sniffing and contract monitoring.
Moreover, a single server contract monitoring
machine can serve a number of CDN replicas which speaks for the practical
server-side deployment of UgoVor.

Finally, beyond detecting subpar service, multilateral micro-auditing for
Internet streaming can improve currently heavily coarse-grained streaming
redirection practices~\cite{CDNbroker:17}, and help advance the existing
performance-agnostic pricing models towards novel and transparent
pay-what-you-experience models. Hence, we discuss
how UgoVor can play an important role in these two directions. 
First, UgoVor
allows CDNs and CDN brokers, currently reliant on aggregate quality
indicators, to leverage UgoVor for fine-grained streaming management. 
Second, UgoVor enables performance-centric valuation and
pricing models that can close the gap between the streaming quality
that clients are paying for on one hand, and actually experiencing on the
other.\footnote{Such novel pricing models do \emph{not} need to 
add uncertainty on how much a user will be charged -- the user still pays a flat fee upfront, yet gets
reimbursed, partially or fully, when the quality of the service is subpar.}

To summarize, our main contributions are the following: \begin{itemize}\itemsep1pt
      \parskip0pt \parsep0pt

    \item We introduce multilateral streaming contracts which offer
      transparent and consistent auditing of streaming sessions by all
      parties involved. % for the first time.

    \item We identify that information asymmetry between clients and servers
      is the key roadblock to the correct and scalable monitoring 
      of streaming contracts, and we
      remove it by introducing a virtual buffer on the server side. 

    \item We implement and release UgoVor, a monitoring system for streaming
      contracts. UgoVor relies on a generic client-server streaming model, and as
      such is compatible with any existing underlying streaming algorithm
      that we are aware of.

    \item We deploy UgoVor in a testbed and,  by
      emulating a Internet-scale real-world streaming trace, we demonstrate that it scales
      and is capable of accounting for each and
      every chunk resolution and rebuffering event. 

  \item We discuss how UgoVor opens the doors to novel, fine-grained
    performance-based streaming management and pricing policies.

\end{itemize}

%

%\section{Motivation}
%\input{section/motivation.tex}

\section{The Design of \tool}
\label{design}
As discussed in the introduction, there are two significant technical issues that the design of \tool
needs to deal with head on. First, \tool targets a setting with asymmetric
participants; the streaming client, e.g., the video player of a browser,
and the server,e.g., a CDN that the content provider employs, have
different views of a streaming session. Specifically, the server knows
when and what video chunks it sends to the client but, due to network
effects, cannot reliably determine \emph{when} the client receives and
actually plays these chunks. Besides the client and the server, the
content provider  should also be able to observe and agree on at least
some of the parameters of the service's quality.  In many cases it may be
in control of the client's player and thus have access to the client's
view of the session but it has no independent access to the server's view.

In response to  this challenge, \tool's contract language focuses on
quality of service parameters that any point in the network can soundly
estimate as long as it has access to the confirmed video chunks the server
sends to the client.  By default the client and the server are such points
but \tool also supports further observers of the service. Their role is to
exchange information with the server and the client and make it
available to third-party auditors or even the content provider when it
can't obtain the information directly from the client. Thus all interested
parties can  share a common view of the quality parameters that are
expressible in \tool's contract language`, by design, agree if the
quality of the service is acceptable or not.

The second challenge is that \tool cannot assume that clients and servers
are honest.  After all they both have very good financial reasons to
``lie''. \tool deals with this issue by discouraging dishonesty through a
careful management of the incentives of all involved parties.  As we
discuss in the introduction, \tool relies on that fact \emph{if the
parties disagree on their view of the service, one of the parties must be
dishonest.} Hence, \tool plays termination of the service against
dishonesty and monitors the service at micro-intervals, dubbed contract
windows, to quickly  minimize any short-term benefits for a dishonest
party. However, this does \emph{not} mean that if, for example, the
user's ISP or home network has a poor connection and cannot achieve a
target video quality, that UgoVor will disallow the user to watch any
videos. UgoVor terminates a session only when one of the parties is
provably dishonest. 

\subsection{\tool Contracts}

 \begin{figure}[tb] 

  \begin{ugovor}[xleftmargin=1pt] 
    { "window"      : 120,
      "resolution"   : [[["720p", 0.5], ["1080p", 1], ["4K", 1]], 
                               [["720p", 0.7], ["1080p", 1], ["4K", 1]], 
                               [["720p", 0.9], ["1080p", 1], ["4K", 1]]],
     "rebuffering"  : [1, 5, 10] } 
  \end{ugovor} \vspace{-2em} \caption{An
 Example Streaming Contract} \label{fig:contract} \end{figure}

The language of \tool contracts is limited but tailored to streaming
quality.  Figure~\ref{fig:contract} demonstrates a typical \tool contract.
In general, \tool contracts are JSON objects with  three key/value pairs:
($i$) The value of the |"window"| key corresponds to the duration in
seconds of the portion of the streaming session that the contract applies
to. Once the window expires, \tool starts checking the contract afresh
with a new window.  ($ii$) The value of |"resolution"| is an ordered list
of lists of pairs. Each pair maps a video resolution to the maximum
percentage of the contract's window that the streaming video can have that
resolution.  Each list of pairs indicates  a contract level.   When \tool
discovers that the service does not live up to the contract at a given
level, for the remainder of the window it starts monitoring the subsequent
level. Outside \tool, the contract parties can agree on a pricing schema
for the different contract levels (obviously, a lower QoS level incurs a lower price), including the price of service that 
fails at all levels.  Back to figure~\ref{fig:contract},
the list
    
    \begin{ugovor}[xleftmargin=1pt] 
      [["720p", 0.5], ["1080p", 1], ["4K", 1]]  
    \end{ugovor} \vspace{-1em} 

    \noindent specifies that the strictest level of the contract requires
    that the video has resolution 720p for at most half of
    each 120 second window of the contract and for the rest of the window
    it can have either resolution 1080p or 4K.  ($iii$) The value of
    |"rebuffering"| constraints is the ordered list of the maximum allowed number of rebuffering
    events per contract level during the contract's window. 

    For simplicity, in the remainder of this section, we focus on
    contracts for video resolution and number of rebuffering events.
    However, in section~\ref{extension}, we extend contracts to include
    rebuffering events' \emph{duration}. 

\subsection{The architecture of \tool}

 \begin{figure}[tb] 

   \includegraphics[width=\columnwidth]{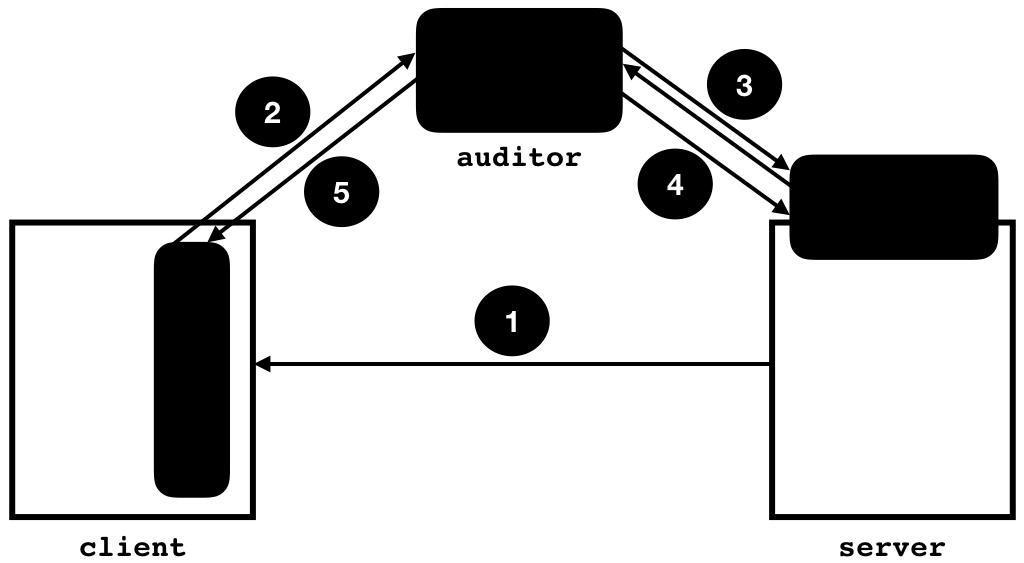} \vspace{-2em}
 \caption{\tool at Work} \label{fig:UgoVor-work} \end{figure}

Figure ~\ref{fig:UgoVor-work} depicts \tool's three components (in black)
and how they cooperate to monitor contracts.

The \emph{client monitor} is part of the client's machine. It's role is to
($i$) intercept messages between the client and server; ($ii$) monitor the
video player of the client's browser to assess the status of the buffer of
the player and the resolution of the video chunks the client receives.
Thus it gathers all the necessary information to check locally if a
contract holds. When it detects a change in the quality of the service
session, it notifies the rest of \tool to
validate the status of the service session with respect to the contract.
In other words, the client monitor is the piece of \tool that  discovers
and signals contract violations.

The \emph{server monitor} confirms or challenges any violations the client
monitor signals. However, to do so, due to the inherent asymmetry
between the client and the server, the server monitor needs to simulate
information that the client monitor directly obtains from the client's
player. In other words, the server monitor maintains a \emph{virtual}
video player buffer.  For that, the video player's configuration comes
with a map from the byte range of the  video file chunks that the server
hosts to their resolution and duration.
%\footnote{This requires the
%\tool-specific pre-processing of the server's video files. However the
%pre-processing can be combined with the creation of the server's
%manifest.} 
For scalability,  the main piece of the server monitor that
performs the core contract-related tasks --- \emph{i.e.} maintaining the
virtual buffer and keeping track of video resolution statistics ---
resides on a machine other than the server.  This piece of the monitor is
also responsible for communicating with the rest of \tool to confirm or
challenge the information the client monitor collects. The server monitor
has a secondary piece that  lives on the server machine and carries out
the lightweight task of sniffing the HTTP(S) messages that the server sends
to the client.  After each message, the sniffer forwards byte range
metadata from the payload of the message to the main piece of the server
monitor to update of the monitor's state.  
       
 The \emph{auditor} is the component of \tool that the client and the
 server monitor communicate with to confirm their views of the quality of
 the service.  If it detects a disagreement between the server and the
 client monitor, 
%or a contract violation, 
it notifies the two monitors to end the service. However, the role of the
auditor goes beyond making sure that all parties are in sync and enforcing
contracts.  It is an independent mechanism that allows the content provider
 to audit the quality of the streaming service.

\subsection{\tool in Action}

To describe the workings of  \tool, herein, we focus on a simple, yet
comprehensive, model of video streaming.  In particular, we assume that
\tool monitors a single live streaming session over a single connection
between a client and a server. We discuss scenarios that go beyond this
model in section~\ref{extension}. Furthermore we assume that both the
client and the server have deployed \tool and that their monitors ``know''
the contract and the address of the auditor.  We explain how we bootstrap
\tool in Appendix~\ref{appendixBS}.  
%However,
Note that the exposition in this section is intentionally agnostic of how
the server decides which data to send when to the client to reflect that
\tool is compatible with any streaming algorithm we are aware of. 

The active role of \tool starts as soon as the server pushes data to the client (step 1 in
Figure~\ref{fig:UgoVor-work}).  The server monitor intercepts the
server's message and updates its virtual buffer to reflect a conservative
approximation of the client's buffer and the resolution related
statistics of the video. 

In parallel, the client monitor  collects information about the service on
its own by tapping directly into the video player's buffer.  In detail,
the client monitor uses the information it collects to detect \emph{events
of interest}: ($i$) a rebuffering event; ($ii$) a change in the resolution
event or; and ($iii$) a contract violation event due to lack of change in
the resolution of the video.  \noindent When it detects one of these
events, it notifies the auditor about the kind of the event and sends
along any relevant information (step 2  in Figure~\ref{fig:UgoVor-work}).
For rebuffering events it sends along  the time point in the video when
rebuffering occurred, dubbed the presentation timestamp of the event.  For
a change in resolution event, it sends the new resolution of the video
together with the presentation timestamp of the resolution change.  For a
contract violation event, it sends the resolution changes in the current
window together with their presentation timestamps. 

The auditor reacts to the receipt of notifications from the client monitor
by asking from the server monitor to send its side of the story (step 3).
If the notification from the client monitor is for a rebuffering or change
of resolution event at some presentation timestamp then the auditor asks
the server monitor to confirm whether an event of the same kind occurred
at that presentation timestamp.  If the notification is for a contract
violation event, the auditor asks the server monitor to confirm the
resolution changes in the current window of the contract and their
presentation timestamps.

In turn, the server monitor  uses its virtual buffer to procure its reply
to the auditor. Specifically, the server monitor updates its buffer as
soon as the server sends a video file to the client. Thus, the virtual
buffer records what file is served at what point in time from the
perspective of the server. Hence, for video resolution, the
server monitor can simply use its buffer and its
configuration to figure out the video quality at particular presentation
times.

For rebuffering information, in addition to the virtual buffer and the
configuration, the server monitor needs to take into account the HTTP(S)
responses (which we call acknowledgments) from the client. In particular,
from the viewpoint of the server monitor, the only way the server monitor
can dispute a rebuffering event is only when the server receives an
acknowledgment for the delivery of the ``missing'' chunk before the end of
the preceding video chunk in the buffer. Put differently, the server
monitor should always confirm to the auditor that there is rebuffering
after the end of chunk $A$, if $t_{A} + length(A) \leq t^{ACK}_{B}$ where
$t_{A}$ is the server timestamp for the outgoing message that carries $A$
and $t^{ACK}_{B}$ the server timestamp for the receipt of the
acknowledgment for the ``missing'' chunk $B$,  while $length(A)$ is the
length of the video playtime of chunk $A$. Note that this condition is
compatible with a worst case scenario; the virtual buffer
of the server assumes that the client's player starts playing chunk A as
soon as the server starts sending it and requires that the next chunk B is 
downloaded and confirmed by the client monitor before the duration of A
eclipses. 

\begin{figure} \includegraphics[width=\columnwidth]{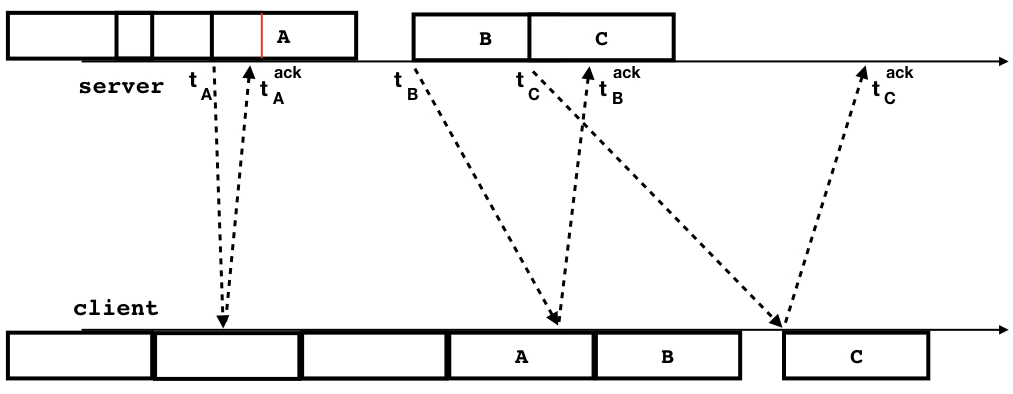} \vspace{-2em}
  \caption{The monitors' views of the video player buffer. The client
  monitor has direct access to the player's buffer. The server monitor
  builds a virtual buffer with the timestamps and the metadata of the
  messages to the client. It detects a rebuffering event when the
acknowledgment for a chunk arrives after the expected end of the preceding
chunk.} \label{fig:buffer} \end{figure}

Figure~\ref{fig:buffer}
depicts how the server monitor populates its buffer side by side with the
view of video player buffer from the perspective of the client monitor.
For chunk $A$ the server monitor knows that it is delivered in time as its
acknowledgment arrives before the estimated end of the preceding chunk
(the red line in the figure). However for chunk B it cannot claim the same
as its acknowledgment arrives after the expected end of chunk A even
though no actual rebuffering took place.  We return to how incentives
minimize the risk that a client can take advantage of such situations in
section~\ref{dishonest}. The important point here is that the virtual
buffer of the server monitor
guarantees that the server detects all real rebufferings such as the one
between chunks B and C.

After the server monitor's reply to the auditor (step 3 in
Figure~\ref{fig:UgoVor-work}), the auditor decides further actions.  If
the two parties agree about a change in an event of interest, the auditor
notifies the monitors (steps 4 and 5) that that they are in sync and the
session should proceed (with a new contract window or a downgraded
contract level if necessary). If the auditor doesn't confirm that the
client and server monitor are in agreement, they interrupt the service.

If all parties are honest, our discussion so far is exhaustive.  For video
resolution, the views of the two monitors are based on the metadata of
HTTP(S) messages. Thus, they cannot have  a
different ``view'' of an event.
For rebuffering, each keeps track of its own buffer,  but as we
describe above, the server's one is always a conservative
approximation of the actual buffer of the client since the two sync-up
based on the client's acknowledgments. Furthermore, the client monitor 
that has access to most precise
information about the client's buffer is the one that detects events of
interest; the server's monitor simply confirms or disputes them.  These
points guarantee the
accurate detection of contract violations. If both parties are honest, as
all the intermediate events of interest
originate from the client and cannot be disputed, a contract violation
detection is always valid.

\subsection{Handling Dishonest Parties}
\label{dishonest}

The correct operation of \tool does not rely on parties behaving honestly.
Given  that the messages monitors receive from another party are indeed
from that other party, which can be guaranteed with standard encryption
techniques set in place as part of \tool's bootstrapping, parties need to
trust  only their own monitor.  First, the parties exchange information to
verify that they are in agreement about some event, not to make a decision
about whether an event has occurred. Hence, they can independently detect
a contract violation. 
%When they disagree, each decides independently that the service should be
%terminated.  
Second, since the parties sync up at short time scales, \tool minimizes
the benefits of dishonesty and, as a result, disincentivizes dishonest
parties all together.

In more detail, if the client monitor reports an event of interest
falsely, then the server monitor disputes it and the session terminates.  Similarly, if the  server monitor falsely
disputes a real event, the session terminates again.  A session
termination never benefits a server as it implies potential loss of
income. As for a client, it can only benefit from the termination if it
comes after a significant portion of the session and if the pricing
schema implies that the client does not have to pay for this portion.
However, ($i$) \tool contracts come with windows whose size the server can
tune to balance this risk; and ($ii$) the result of contract checking
depends only on intermediate mutually agreed-upon events that are
validated as they occur. Finally, the auditor has also no incentive to either
report a false disagreement or ignore an event of interest from the client
monitor; it represents the content provider whose interests demand that
the streaming session continues as long as the client and the server agree
its quality is acceptable.

There is one additional way a client monitor can be dishonest: delayed
acknowledgments.  With delayed acknowledgments, the client may try to to
trick the server monitor to confirm a false rebuffering event and cause
the contract to switch to a lower contract level (with a lower price).
However, (i) this is only for the short duration of the current contract
window; and (ii) if the client does not also adapt its requests for chunks
to a bit rate compatible with the congestion implied by the delay of the
acknowledgments, the server monitor can detect it, report it to the
auditor and terminate the session. 

As a final comment, \tool is also resilient against collusion between  the
auditor and  one of the other two parties, the client or the server. Specifically, from the perspective of the 
non-colluding party the situation is no different that dealing with a dishonest party.
Of course if the client and the server collude then the auditor is in the
dark about the performance of the streaming session. In practice, though,
we anticipate that content providers will deploy their own client monitors
alongside those of end-users.

%\section{Implementation}
%\input{section/implementation.tex}

\section{Evaluation}
\label{eval}

%In this section, we evaluate the performance of a proof-of-concept
%implementation of \tool. To do so,  we exploit a 10-month-long data set from a streaming
%service.  Below, we first explain the properties of the data set and
%analyze it in order to assess its effects on Ugovor's sclability.  
%We then describe our experimental set up for assessing \tool's using the
%data set. Finally, we explore UgoVor's application in two
%scenarios: fine-grained stream redirection control and performance-based
%pricing.

\subsection{Live Streaming Data Analysis}

We were provided access to a streaming data set collected from Puffer, a
Live TV streaming service~\cite{puffer19}. The service streams U.S. TV
stations affiliated with the NBC, CBS, ABC, PBS, Fox, and CW networks.
Clients can watch the programs in their browser, \emph{i.e.}, Chrome,
Firefox, and Edge browsers. All the clients were served from a single
server with 10 Gbps connectivity in a well-provisioned datacenter.  Traces
from this Live TV service were captured over a 10-month period and include
a total of 439,900 streaming sessions with clients from 1332 unique ASes.
The traces amount to 25.7 years worth of streaming data with sessions on
mobile as well as desktop environments.  

\emph{Effects on UgoVor}: While it is unknown to what extent Puffer's
results (data center streaming over a wide-area network to clients)
generalize to typical paths between a user on an access network and a
nearby CDN server~\cite{puffer19}, it still provides insights from a
large-scale streaming experiment. More importantly, it does not prevent us
in any way from accurately evaluating UgoVor's performance, while
retaining the invaluable real-world effects existent in the data set.

\begin{figure*}[h]
  \centering
  \begin{minipage}[t]{0.25\textwidth}
    \centering
    \captionsetup{justification=centering}
    \includegraphics[width=1.7in]{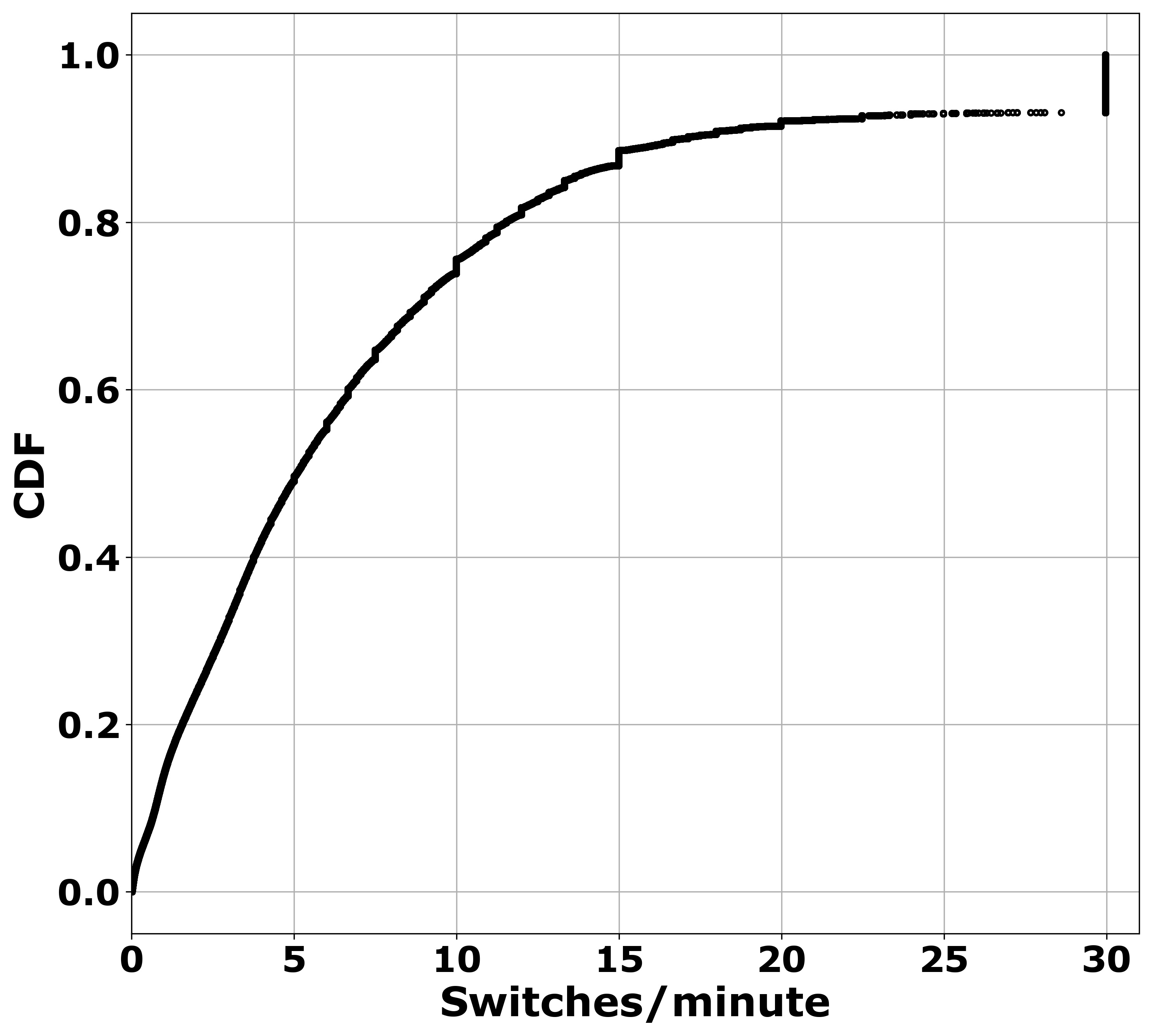}
%    \caption{Distribution of the number of switches in resolution per minute within a session}
%    \label{fig:spm}
  \end{minipage}\hfill
  \begin{minipage}[t]{0.25\textwidth}
    \centering
    \captionsetup{justification=centering}
    \includegraphics[width=1.7in]{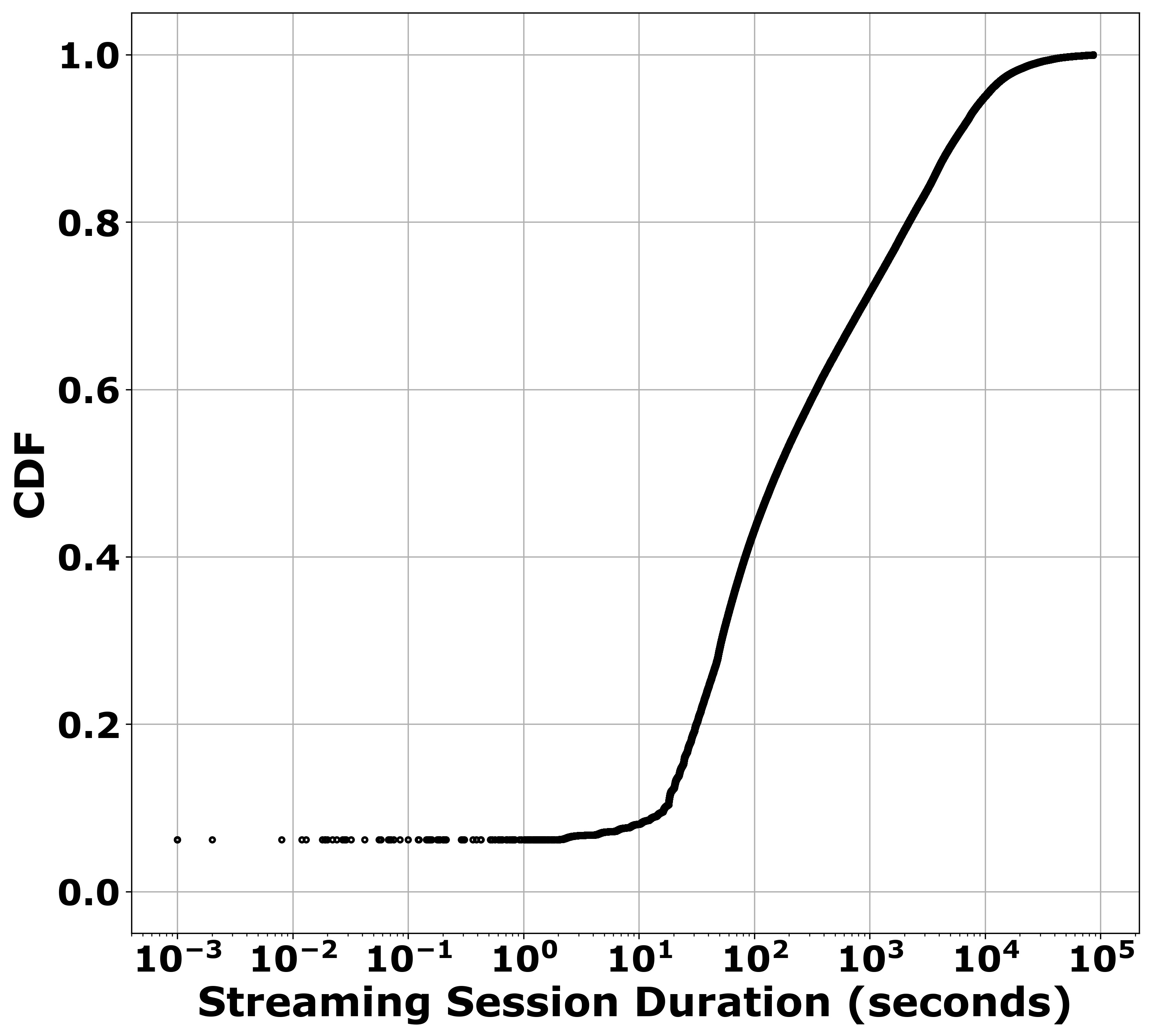}
%    \caption{Distribution showing the Duration of Streaming Sessions}
%    \label{fig:total_dur}
  \end{minipage}\hfill
  \begin{minipage}[t]{0.25\textwidth}
    \centering
    \captionsetup{justification=centering}
    \includegraphics[width=1.7in]{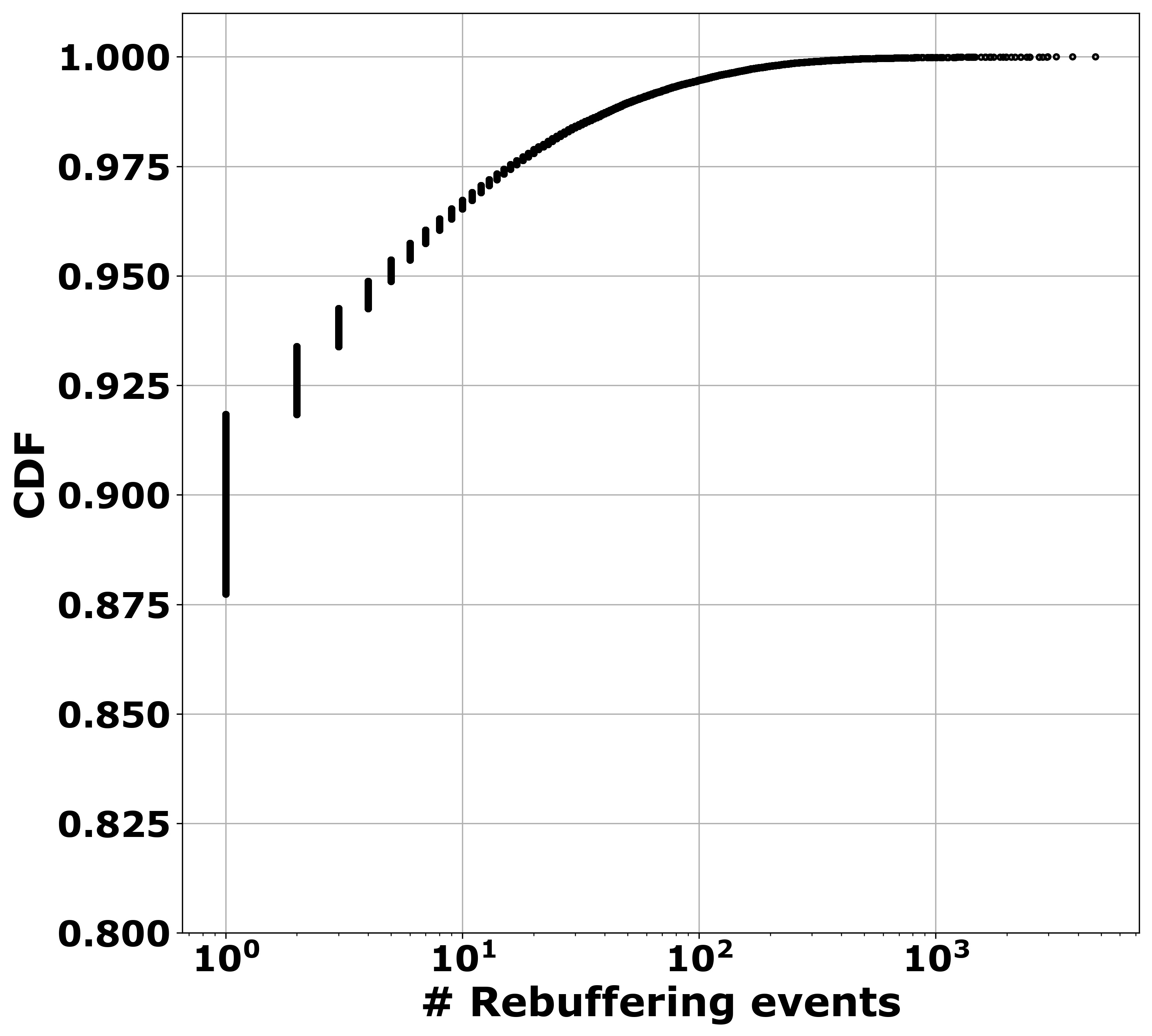}
%    \caption{Distribution showing the number of rebuffering events}
%    \label{fig:rebuf_dur}
  \end{minipage}\hfill
  \begin{minipage}[t]{0.25\textwidth}
    \centering
    \captionsetup{justification=centering}
    \includegraphics[width=1.7in]{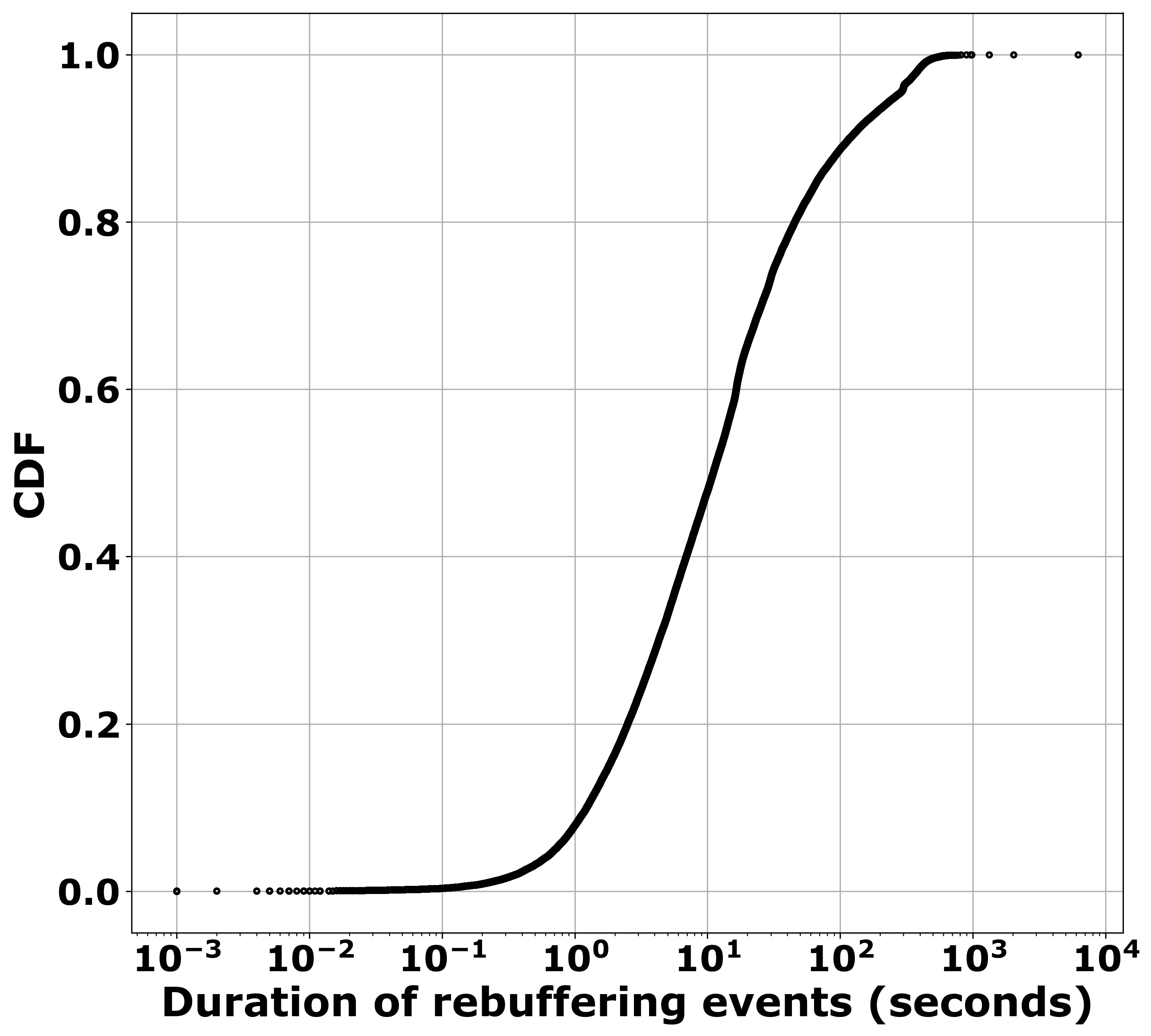}
%    \caption{Distribution showing the duration of rebuffering events}
%    \label{fig:rebuf_dur}
  \end{minipage}\hfill
    \caption{Aggregate results over the entire data set, showing (from left to right): (a) CDF of the number of video resolution switches per minute within a session, (b) CDF of the duration of streaming sessions, (c) CDF of the number of rebuffering events per streaming session, and (d) CDF of the duration of rebuffering events.}
    \label{fig:agg}
\end{figure*}

\subsubsection{Aggregate-Level Results} 

%\begin{table}
%\centering
%\begin{tabular}{cc}
%Resolution & Percentage  \\
%240        & 5.51        \\
%360        & 2.91        \\
%480        & 4.52        \\
%720        & 41.11       \\
%1080       & 45.95      
%\end{tabular}
%\caption{Fraction of each quality level in the streaming data set}
%\label{table1}
%\end{table}

We first analyze 
%Table~\ref{table1} shows 
the share of each streaming quality level in all streaming sessions put together. 
Independent from the streaming quality resolution, each streaming chunk accounts for 2 seconds of the streamed video.
We find that resolution levels 720 (also known as HD Ready) and 1080 (also called Full HD), collectively account for 
87.06\% of all chunks streamed. We refer to this group as High Definition or HD resolutions. 
%sum up to 87.06\% of all chunks streamed. 
The remaining resolution of chunks, \emph{i.e.}, levels 480, 360, and 240, which we call Standard Definition or SD resolutions, collectively 
account to 12.94\%. It is clear that while HD chunks dominate the streams, there is still a substantial amount of chunks that are below HD quality. Moreover, quality switches 
among any of the five resolution levels can happen across the duration of a stream. 
%We quantify this next.

%https://www.digitalcitizen.life/what-screen-resolution-or-aspect-ratio-what-do-720p-1080i-1080p-mean
%720p = 1280 x 720 - is usually known as HD or "HD Ready" resolution
%1080p = 1920 x 1080 - is usually known as FHD or "Full HD" resolution
%1440p = 2560 x 1440 - is commonly known as QHD or Quad HD resolution, and it is typically seen on gaming monitors and on high-end smartphones. 1440p is four times the resolution of 720p HD or "HD ready." To make things even more confusing, many premium smartphones feature a so-called 2960x1440 Quad HD+ resolution, which still fits into 1440p.
%4K or 2160p = 3840 x 2160 - is commonly known as 4K, UHD or Ultra HD resolution. It is a huge display resolution, and it is found on premium TVs and computer monitors. 2160p is called 4K because the width is close to 4000 pixels. In other words, it offers four times the pixels of 1080p FHD or "Full HD."
%8K or 4320p = 7680 x 4320 - is known as 8K and it offers 16 times more pixels than the regular 1080p FHD or "Full HD" resolution. For now, you see 8K only on expensive TVs from Samsung and LG. However, you can test whether your computer can render such a large amount of data using this 8K video sample:

\emph{Effects on UgoVor}: 
In terms of \tool, an important parameter is the number of resolution
changes over time.  Each such event requires a communication between the
client and server monitors and auditors.  Figure~\ref{fig:agg}(a)
depicts the distribution of the number of resolution changes per minute
within a session.  We can see that the median number of resolution changes
per minute is about 5. Necessarily, this includes both resolution upgrades
and downgrades. The figure also shows that around 5\% of streams
experience exactly 30 changes per minute. For these flows, every
consecutive chunk, worth 2 seconds of viewing time, has a change in the
resolution. Such behavior induces the most load on UgoVor,
however, the percent of such streams is rather small.

Next, we focus on the reminder of Figure~\ref{fig:agg}, and analyze
the distribution of ($i$) the duration of the flows in the data set
(Figure~\ref{fig:agg}(b)), ($ii$) the number of rebuffering events
per streaming session (Figure~\ref{fig:agg}(c)), as well as ($iii$)
the duration of rebuffering events (Figure~\ref{fig:agg}(d)).
Figure~\ref{fig:agg}(b) shows that the median streaming duration is
155.61 seconds, while the longest stream lasts for as long as 27.7 hours.
Regarding the rebuffering statistics, Figure~\ref{fig:agg}(c) shows
that approximately 12.5\% of streams experience rebuffering events. In
particular, approximately 5\% of the streaming sessions experience a
single rebuffering event, while the remaining sessions experience two or
more. Figure~\ref{fig:agg}(d) shows that the rebuffering events are
quite severe in terms of duration, \emph{i.e.}, 95\% of them are longer
than 1 second, and approximately 57\% of rebuffering events are longer
than 10 seconds.

We provide AS-level analysis
%, and its impact on \tool, 
in Appendix~\ref{appendixAS}.

\subsection{Evaluating \tool with Emulation Streaming in a Testbed}

Here, we first explain how we utilize the Live TV streaming data set
analyzed above to evaluate UgoVor. In particular, the data traces contain
entries for each video client session including the times at which video
chunks were sent and acknowledged, resolution of the video chunks, as well
as the presentation timestamps and the size of the chunks. Additionally,
the traces also include client-side events such as video-player
interactions, rebuffering events, and presentation buffer health at
regular time intervals.  While UgoVor is completely independent from the
type of the endpoint streaming algorithm utilized, next we provide the
necessary background on Puffer, because it fundamentally affects our
ability to effectively emulate real-world streams in a testbed.

Puffer is an ML system that uses \emph{no} explicit application-level
requests from the browser-receiver. Instead, it sends data chunks to the receiver
via HTTP and utilizes its own measurements to decide when to send the
chunks. In particular, it utilizes the sizes of chunks, transmission
time of past chunks, TCP statistics such as current congestion window
size, the unacknowledged packets in flight, the Linux RTT estimate after
smoothing, the minimum RTT as calculated by TCP from Linux, and Linux TCP
estimated throughput.  As explained above, this information is recorded
on the server side along with the packets' timestamps. In our testbed, we
use the information to send (dummy) data to the clients at the
appropriate intervals.  Upon receiving the first chunk, the client
initializes its playhead and fires events based on the state of its receiver.  These
events are used to alarm the auditor, when needed.  While no content is
actually played at the receiver, given that we send dummy data from the
server, the receiver buffer state is accurately reproduced.

\emph{Effects on Ugovor}: Puffer is a "perfect" endpoint streaming
algorithm for testbed emulation, such as for UgoVor's evaluation, because it
uses no application-level feedback. This enables us to send the data at
exactly the same times as in reality. Hence, the chunk
inter-departure times and the chunk resolution levels are decided based on
the real data, \emph{i.e.}, based on the network environment measured by the real
streaming server.  On the receiver side, necessarily, we are able to
perfectly reconstruct the resolution of chunks. However, we are unable to
perfectly replicate the length of rebuffering events, because the chunk
inter-arrival times may not be exactly the same as in reality. Still,
%given that UgoVor counts only the \emph{number} of rebuffering events,
%this is not a problem. We 
we check and confirm that the receivers capture
\emph{all} rebuffering events from the real trace.

{\bf Testbed.} Our evaluation testbed consists of 4 physical machines equipped with 16GB
Memory and 8-core Xeon Silver 4110 processors while equipped with a 1 Gbps
Ethernet connection. Each machine is dedicated for a singular role such
as of the upstream video server, the auditor, the server monitor, and
finally to spawn a pool of clients along with their client monitors. The
client monitor is a python application that inspects the data structures of it's video
player. The application also communicates with the auditor through a well-known TCP port.  The
server monitor and the auditor are multi-process python applications (one
process per client session) that communicate with each other over TCP.
Finally, in the testbed, latencies are artificially induced to match that
from the Puffer data.

%In our experiments, 
To obtain a 95\% confidence level with a
margin of error of 0.05, we randomly select 384 streams from the trace
(see Appendix~\ref{appendix} for details) and emulate them in the testbed.

%FIXME, talk about sampling, how many flows do we sample and why. Explain the statistical logic. 

\subsection{UgoVor's Accuracy}
\label{subsec:accuracy}

The correctness of \tool necessitates that it
enforces contracts accurately. To evaluate \tool's correctness, we 
utilize a contract that matches the average quality  (average resolution
and rebuffering events) of the streams in the
dataset. Figure \ref{fig:best_fit} displays this average contract (top). 

Our experiment proceeds as follows. We first log the events as
experienced  by the clients for a sample of streaming sessions chosen from
the dataset to represent sessions on the busiest day of the server in
terms of data transmitted.  Using the logged events from each session,
we first calculate if the sessions should satisfy our average
contract. Subsequently, we deploy \tool on the same sample of streaming
sessions and verify whether \tool's decision about whether a session
meets its contract matches our independent calculation.

We observe that \tool enforces contracts accurately. In particular, 
the client monitor observes the same events for each session as the 
events we log before deploying \tool.
Moreover, there are no session terminations as result of disagreements 
between server and client monitors. In other words we confirm that 
our honest monitors  have consistent views of the quality of each session
and that quality matches the expected one from our dataset.

This result grants further discussion. Indeed, how is it possible that not a single
session was terminated by UgoVor? All this despite the potentially "weird" timing issues caused by end-to-end 
Internet latency or jitter that can cause the server side to over count potential rebuffering events,
given that it is utilizing a virtual buffer. The fundamental reason for the high accuracy 
is because in our design \emph{the client}, which has the perfect knowledge about rebuffering, exclusively 
raises contract violation challenges. The server's virtual buffer, on the other side, is driven by the acknowledgments
from the client, which are strictly conservative estimates of the times when the chunks are played at the client.
Hence, the server is capable of \emph{deterministically} confirming all actual contract violations
raised by the client.

\subsubsection{UgoVor's Scalability}

\begin{figure*}[h]
  \centering
  \begin{subfigure}[t]{0.25\textwidth}
    \centering
    \captionsetup{justification=centering}
    \includegraphics[width=1.7in]{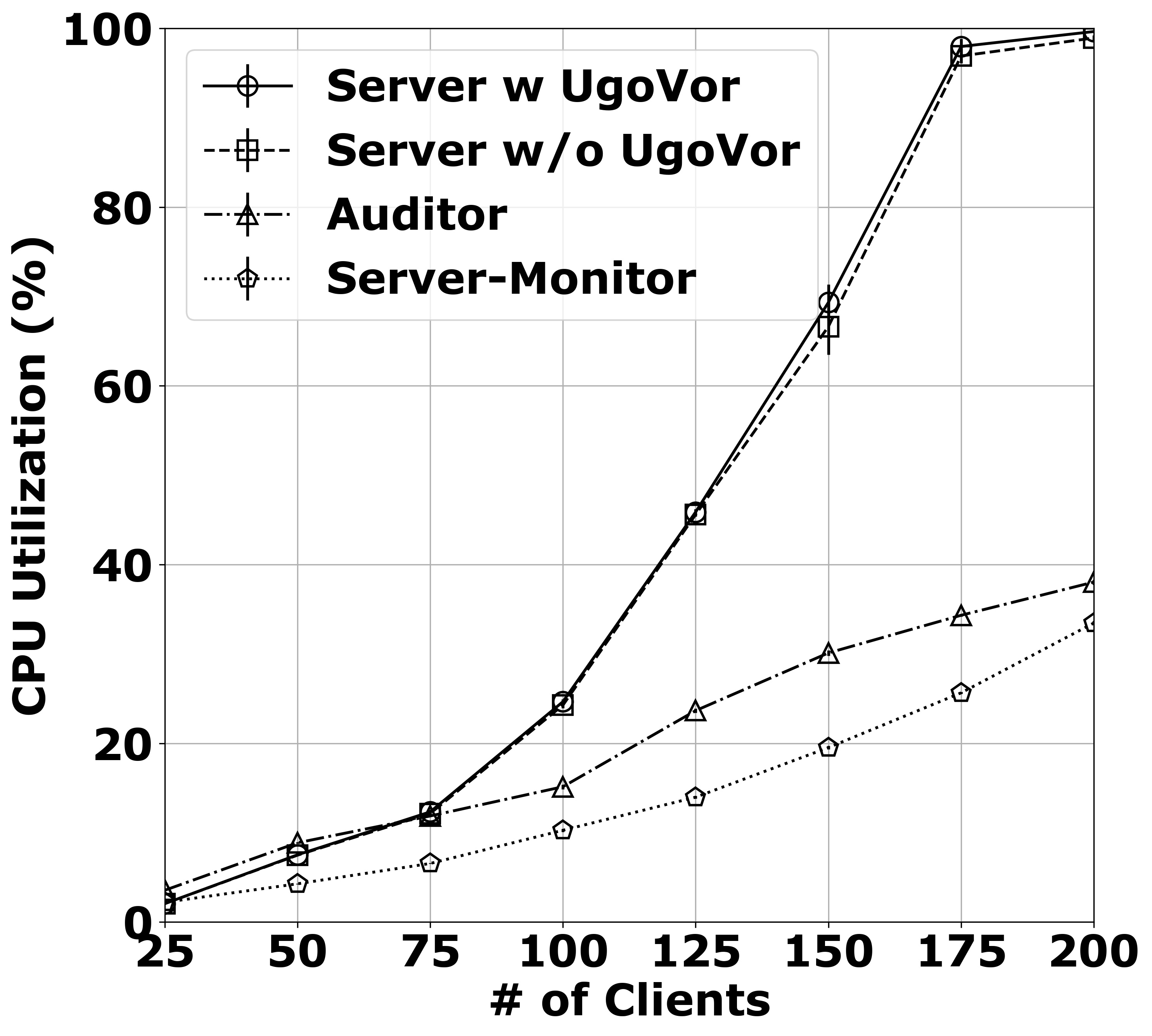}
    %\caption{}
    \label{fig:cpu}
  \end{subfigure}\hfill
  \begin{subfigure}[t]{0.25\textwidth}
    \centering
    \captionsetup{justification=centering}
    \includegraphics[width=1.7in]{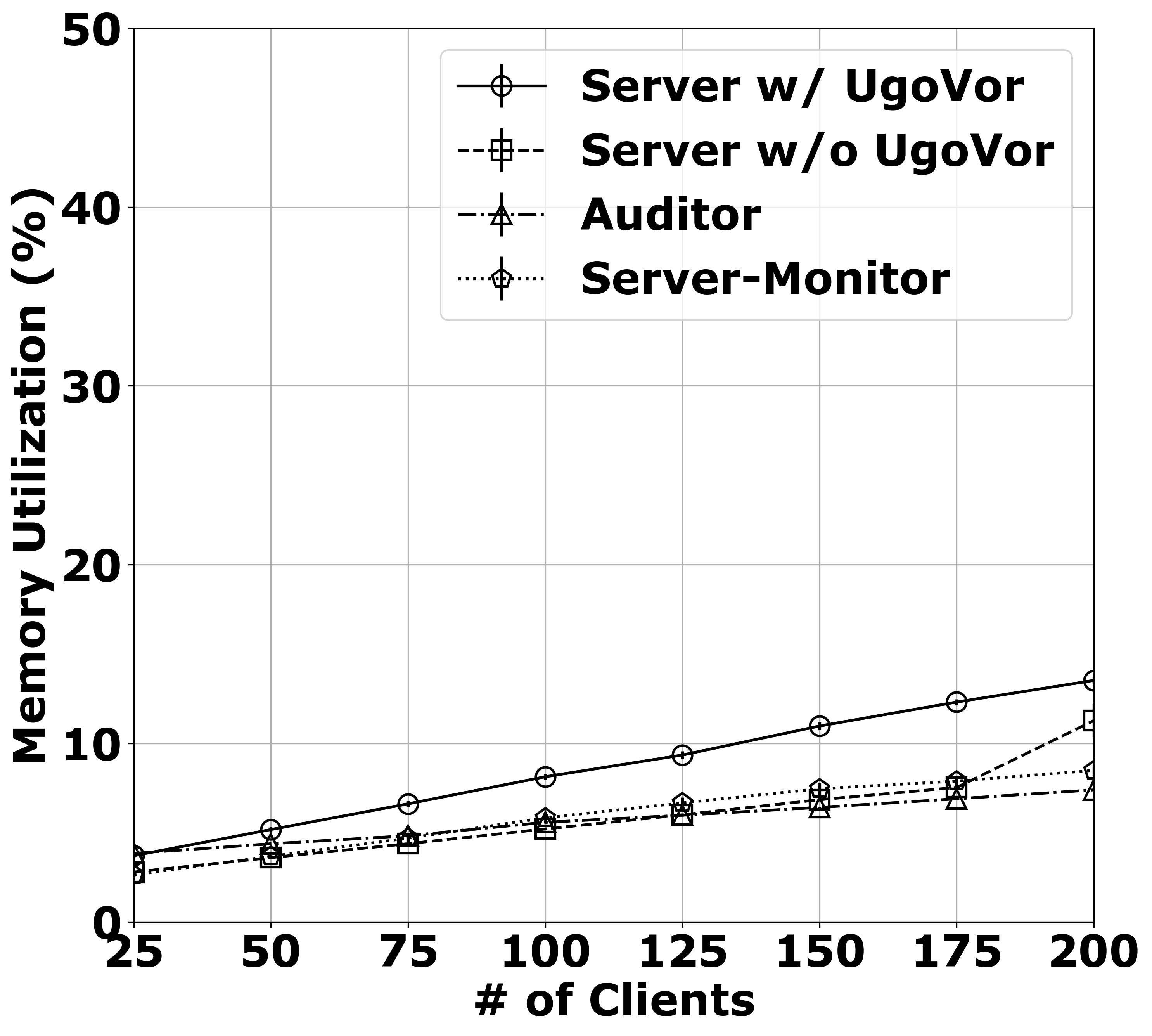}
    %\caption{}
    \label{fig:mem}
  \end{subfigure}\hfill
  \begin{subfigure}[t]{0.25\textwidth}
    \centering
    \captionsetup{justification=centering}
    \includegraphics[width=1.7in]{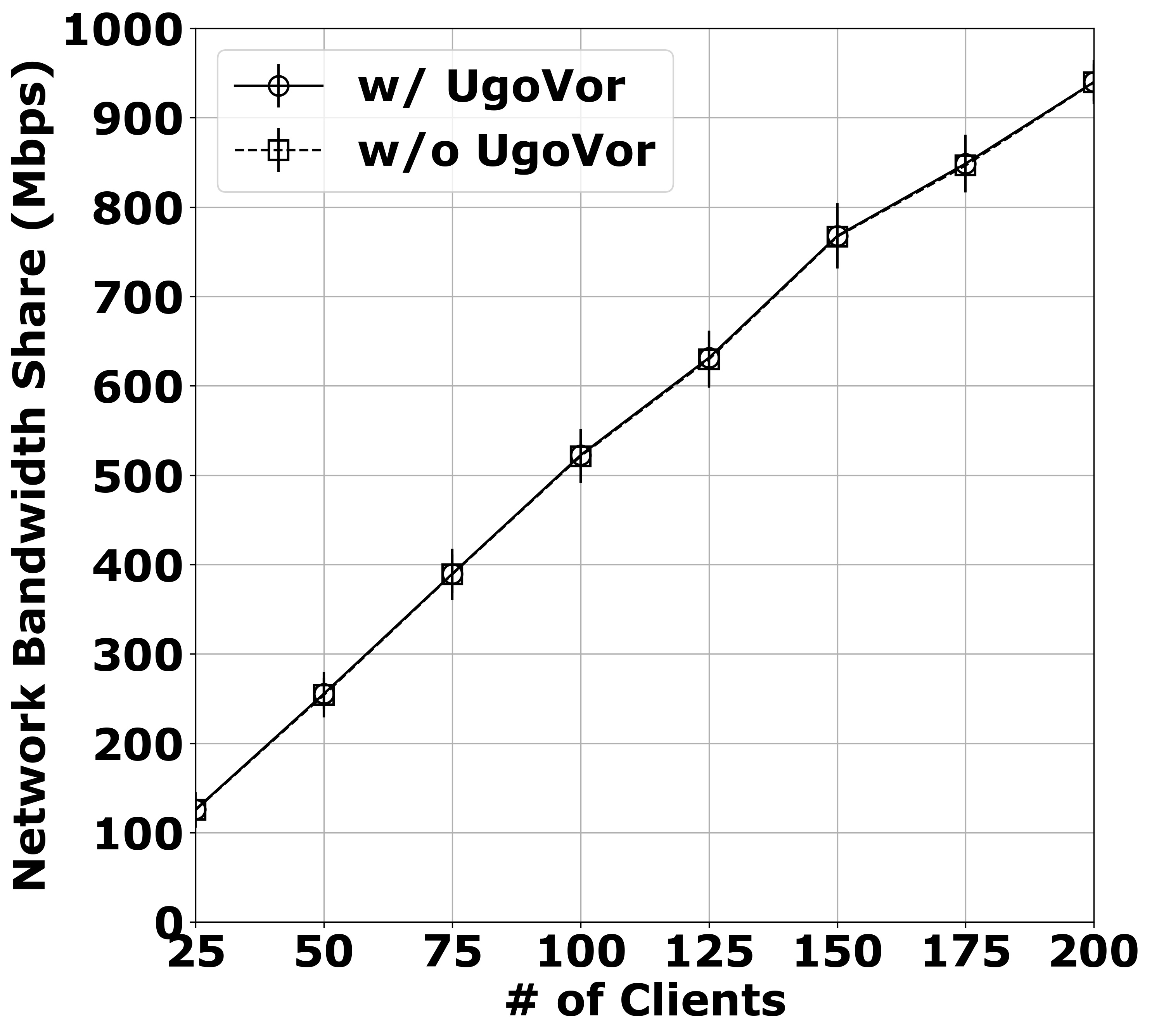}
    %\caption{}
    \label{fig:net}
  \end{subfigure}\hfill
  \begin{subfigure}[t]{0.25\textwidth}
    \centering
    \captionsetup{justification=centering}
    \includegraphics[width=1.7in]{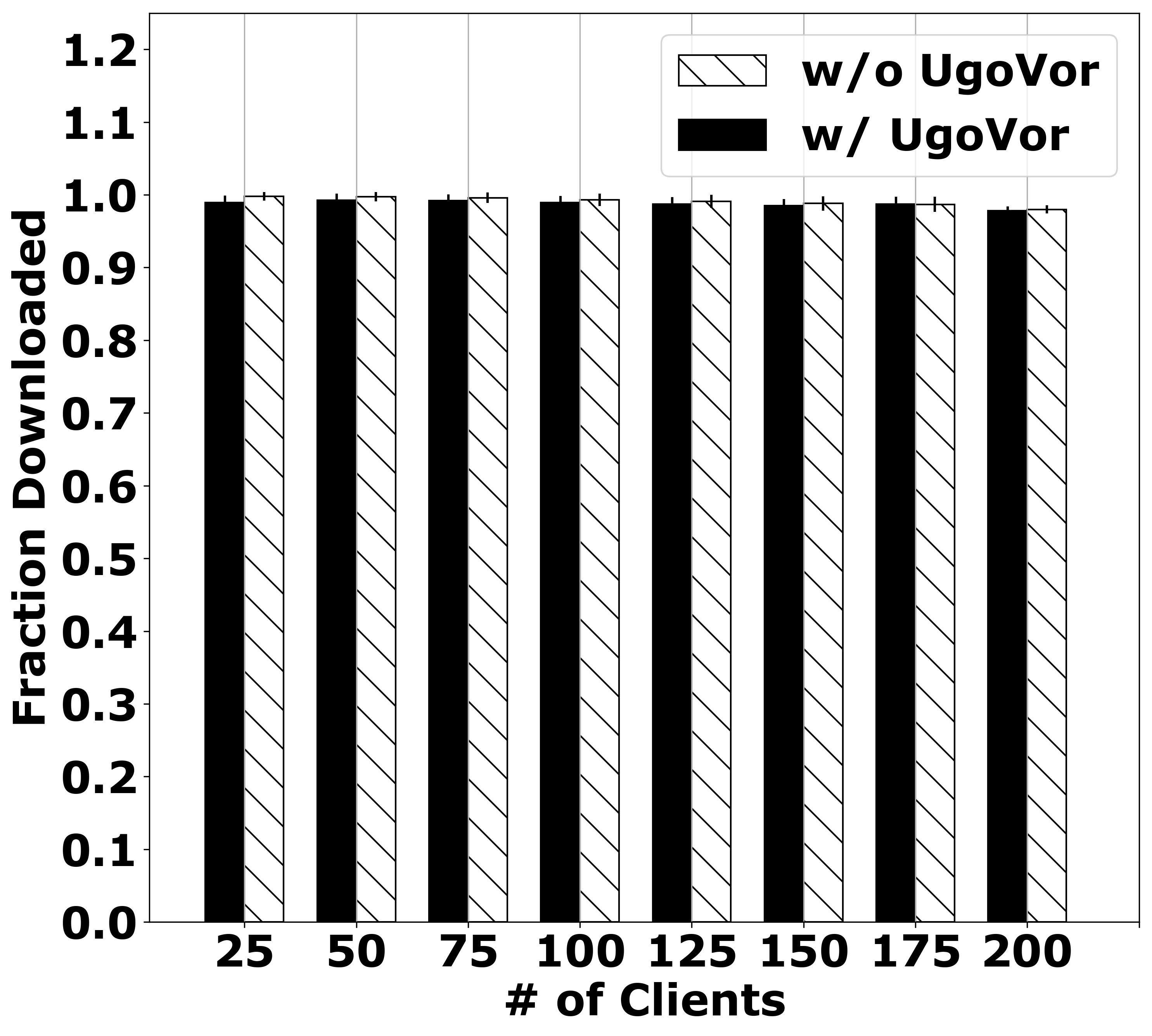}
    %\caption{}
    \label{fig:frac}
  \end{subfigure}
  \caption{(a) Average CPU utilization with varying number of clients, (b)
  Average RAM
  utilization with varying number of clients, (c) Average Server network bandwidth share with varying number of clients, 
  (d) Fraction of expected streaming bytes received by a client in 120 seconds with
  varying number of clients.} 
%connected to the server.}
\label{fig:aggregate}
\end{figure*}

We look into
four aspects of the performance of the testbed and how they are affected
by \tool: ($i$) the maximum number of video streaming clients the upstream
server can support; ($ii$) the server's CPU utilization; ($iii$) its memory consumption and
($iv$) the performance of server/client network connections.

\noindent
\emph{Maximum number of clients.} To understand  the baseline limitations
of the upstream server, 
we ran it without \tool for a varying number of clients from our
testbed and their corresponding flows (the baseline phase).  In particular, we progressively
increased the number of clients until we reached the full capacity 
of the server. Subsequently, we
repeated the same process with \tool deployed for the
server and all clients in a worst case scenario for \tool (the \tool
phase). In particular, ($i$) for all streams
 \tool monitors  the contract 
that accepts the maximum number of rebuffering events in our data and
allows streams of any quality and ($ii$) all streams share the same server
monitor and auditor. Since by default every rebuffering and
resolution change leads to communication between the monitors and
the auditor, and by construction of the contract, there are no contract violations, 
in this setting \tool incurs the  maximum effect possible on the testbed. The conclusion of
the experiment is that despite \tool working at full tilt in the \tool
phase of our experiment, the upstream server can support the same number of
clients (200) as without \tool.

\emph{Server CPU Utilization:} We define CPU utilization as the 
ratio of the total time spent in the CPU to the maximum
available CPU time over 120s of video streaming. To determine the effect of \tool on the CPU
utilization of the server, similar to above, we measured the performance of the
testbed in  the baseline and the \tool phase.  
In detail, for both phases, we selected random samples of sizes
between 25 and 200 flows from the pool of flows (384 random samples per size --- see
appendix~\ref{appendix} for details).
Figure~\ref{fig:aggregate}(a)
%\ref{fig:cpu} 
shows the
mean CPU utilization for each sample size. In the baseline phase the server
reaches over 95\% CPU utilization when supporting
 samples of size 175 and above. In the  \tool phase, 
the CPU utilization closely follows that of the baseline phase for all sample
sizes. The most significant increase in CPU utilization between
the two phases is 4.05\%. This should not come as a surprise; the only additional
function of the video server in the \tool phase is sniffing and forwarding
to the server monitor
metadata about the video chunks served to the clients.
As further evidence of the CPU load from the use of \tool, Figure~\ref{fig:aggregate}(a) 
%\ref{fig:cpu} 
also shows that maximum average CPU utilization for the auditor and the server-monitor 
 reaches is less than 39\% and occurs for the maximum sample size (200).

\emph{Server Memory Usage:} 
For memory usage, we utilize the same experimental setup as for CPU
utilization. That is we create random samples of varying sizes from the 
pool of streams and we record the accumulative RAM usage of the server for 
all flows in a sample.
Figure~\ref{fig:aggregate}(b)
%\ref{fig:mem} 
shows the mean RAM usage for each sample size. 
While \tool does result in increase in RAM usage for the
server, the difference in RAM usage between the \tool and the baseline
phase of our experiment remains significantly below 15\% even
for samples of size 200. As for the memory usage for the auditor and the
server monitor, \tool causes use of less than 10\% of the available
RAM memory of the machines that run the auditor and the server monitor. 
This modest memory consumption is due to the fact that contract
related data and operations are implemented as in-memory operations.

\emph{Network Performance:} 
Again to measure the effect of \tool on network performance we compare the
performance of the testbed in the baseline and the \tool phase for samples
of flows of different sizes from the pool. First, we measure the average
bandwidth share of the server for the duration of all the flows in each
sample.
Figure~\ref{fig:aggregate}(c) 
%\ref{fig:net} 
shows these measurements as an average per sample
size for both the baseline and the \tool phase of the experiment. 
The two plots are extremely close and the worst case penalty due to 
\tool is 0.51\%. After all, the \tool-related queries that the monitors and the
auditor exchange are intentionally
compact so that they can be accommodated alongside regular
application-related traffic on the network.  
As a second piece of evidence for \tool's cost in terms of network
performance, Figure~\ref{fig:aggregate}(d)
%\ref{fig:frac}
shows the average fraction of streaming related bytes
 over the overall bytes (averaged by
sample size) each client in the sample is expected to receive in 120s of
 video streaming according to the data set.
 Again the baseline phase results are practically indistinguishable
from the \tool phase results.  

%Overall our experiments show that \tool causes acceptable overheads
%in CPU and memory consumption. More importantly \tool does not add significant bandwidth
%load on the network.

\subsection{UgoVor's Applications}
\label{subsec:apps}

%% big idea: applications section
%% Compare status quo of connection mgmt and pricing policies vs if we had fine grained ctcs

%% 2 sections:
%% 1. Existing policies inherently leave behind/harm some customers bc they're coarse.
%%    Fine grained better for everyone.
%%    And make extra point that to satisfy everyone with coarse, must have absurdly permissive policy
%% 2. Existing policies overcharge bc they are oblivious to actual quality delivered
%%    - frame as how much they are LOSING for using status quo
%%    - another?

%UgoVor's fine-grained approach benefits both providers and consumers of
%streaming services by enabling accurate, real-time quality monitoring per
%streaming session.  The coarse-grained nature of prevailing approaches for
%policy management and pricing leave significant portions of clients
%unhappy and overpaying.  With session-level policy monitoring, however,
%connections can be managed and priced optimally for every client.  We
%demonstrate these points with two illustrations of how UgoVor's monitoring
%capabilities would change the streaming experience of clients in the
%Puffer data set.

\subsubsection{Fine-grained Stream Control}

\begin{figure*}
    \begin{minipage}{.4\textwidth}
    \centering
    % include first image
    \includegraphics[width=1.7in, scale=0.5]{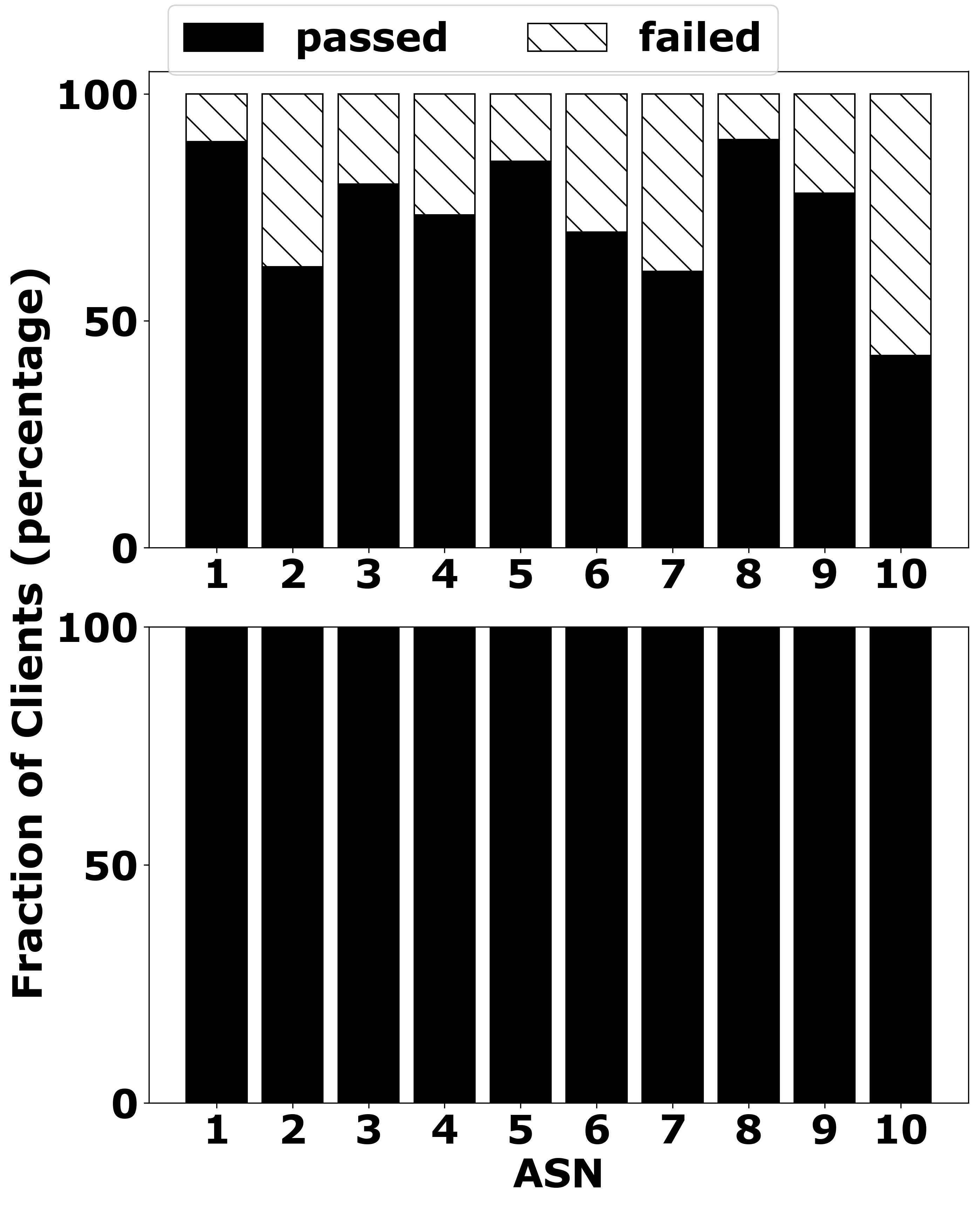}
    \end{minipage}
    \begin{minipage}{.5\textwidth}
    \begin{ugovor}[xleftmargin=1pt]
      { # Average contract
        "window"       : 120,
        "resolution"   : [[["240p", 0.09], ["360p", 0.03], ["480p", 0.08],
                                   ["720p", 0.80], ["1080p", 1]]]                                   
        "rebuffering" : [0]
      }
    \end{ugovor}
    \begin{ugovor}[xleftmargin=1pt]
      { # The most restrictive contract satisfied by all clients
        "window"       : 120,
        "resolution"   : [[["240p", 1], ["360p", 1], ["480p", 1],
                                    ["720p", 1], ["1080p", 1]]],
        "rebuffering" : [33]
      }
    \end{ugovor}
    \end{minipage}
    \caption{Fraction of clients satisfying their respective contracts.}
    \label{fig:best_fit}
\end{figure*}

%By enabling stream control management on the level of individual
%connections, \tool personalizes stream control policy to optimize the
%experience of every client.

Streaming connections are typically managed in the aggregate, with
policies often managing regional groups of clients. 
%called an AS.
In particular, it is well documented that CDN brokers often redirect
entire groups of users from one CDN to another, midstream, and typical
policies are based on aggregate quality indicators~\cite{CDNbroker:17}. 
%from the aggregate~\cite{CDNbroker:17}.  all the clients in the
%group~\cite{CDNbroker:17}.  once the aggregate network measurements or
%policies are triggered~\cite{CDNbroker:17}.  All of the clients within an
%AS stream data from the same content host (TODO: right term?), and
%typical policies transfer entire ASs from one host to another based on
%aggregate quality indicators from all the clients in the AS.
For instance, even if a subset of clients in a group, \emph{e.g.}, an AS,
experience service below some quality threshold, then all of the clients
within that group will switch to use a new content host at another CDN.
%For instance, if 50\% of clients in a group, \emph{e.g.}, an AS,
%experience service below some quality threshold, then all of the clients
%within that group will switch to use a new content host at another CDN.
%
Hence, there are inevitably some clients within that AS who experience
acceptable service but are switched to a new CDN anyway, and the CDN which
served the clients before the switch is severely
punished~\cite{CDNbroker:17}.
%and there are no guarantees that the service will remain acceptable after
%the switch.

Figure \ref{fig:best_fit} illustrates that in the Puffer data, applying
such a policy at the AS level, would force a significant proportion of
clients to experience unnecessary redirections and possible service
disruptions due to aggregate control policies.  Each row in the figure
illustrates the proportion of clients satisfying a policy within each of the 
10 ASes in the data over a period of time.  The policy in the first row
represents an example contract (shown on the right) that specifies a
reasonable quality expectation for every session.  This contract was
computed based on the average quality experienced by the streams in the
data set.  On the left, the bar chart shows the proportion of clients that
do or do not satisfy the quality contract.  The chart demonstrates that
within each AS there can be significant differences in clients' quality
experience, so deciding to switch the entire group to a new content host is
unnecessary 
%often unnecessarily degrades the experience of 
for many users.  For instance, one might consider switching AS number 10
because a large portion of its clients experience poor quality, yet doing
so also unnecessarily redirects 
%interrupts the service of 
the other 40\% of its clients.

Such casualties are ultimately unavoidable for \emph{any} reasonable
aggregate quality policy.  An aggregate policy that does not switch any
clients unnecessarily must be so permissive as to be useless.  To
illustrate this, the bottom contract of Figure~\ref{fig:best_fit} shows the
most restrictive contract that is satisfied by every session in the Puffer
data.  This contract allows any resolution whatsoever and up to 33
rebuffering events every two minutes, which is not a useful policy for
describing acceptable content quality.  Managing all of the clients within
an AS in the aggregate with any reasonable policy is therefore bound to
unnecessarily redirect, potentially disturb the experience of customers
enjoying satisfactory content quality, and severely punish a CDN.

Hence, per-session stream-control enables fine-grained policies that
prevent unnecessary redirections.
%client service disruption.
With UgoVor, individual clients experiencing unacceptable service quality
can be detected and switched to new content hosts without affecting their
peers.  Furthermore, the Puffer data implies
%provides evidence 
that such a strategy would likely improve service for a significant
portion of users in real streaming applications.

\subsubsection{Using UgoVor to Understand Session Value}

%% \begin{figure*}
%%   \begin{subfigure}{.30\textwidth}
%%     \centering
%%     % include first image
%%     \includegraphics[width=0.9\textwidth]{figures/quality.png}
%%     \caption{Client quality experience often falls below HD}
%%     \label{fig:macro-overpay-no-buffering}
%%   \end{subfigure}
%%   \begin{subfigure}{.30\textwidth}
%%     \centering
%%     % include second image
%%     \includegraphics[width=0.9\textwidth]{figures/pricing.png}
%%     \caption{Clients over pay with quality-oblivious pricing}
%%     \label{fig:macro-overpay-2-rebuf}
%%   \end{subfigure}
%%   \begin{subfigure}{.30\textwidth}
%%     \centering
%%     % include third image
%%     \includegraphics[width=0.9\textwidth]{figures/pricing-rebuf.png}
%%     \caption{Considering rebuffering, clients over pay by even more}
%%     \label{fig:macro-overpay-1-rebuf}
%%   \end{subfigure}
%%   \caption{Streaming Price Disparity}
%%   \label{fig:pricing}
%% \end{figure*}
\begin{figure}
    \centering
    \includegraphics[width=0.3\textwidth]{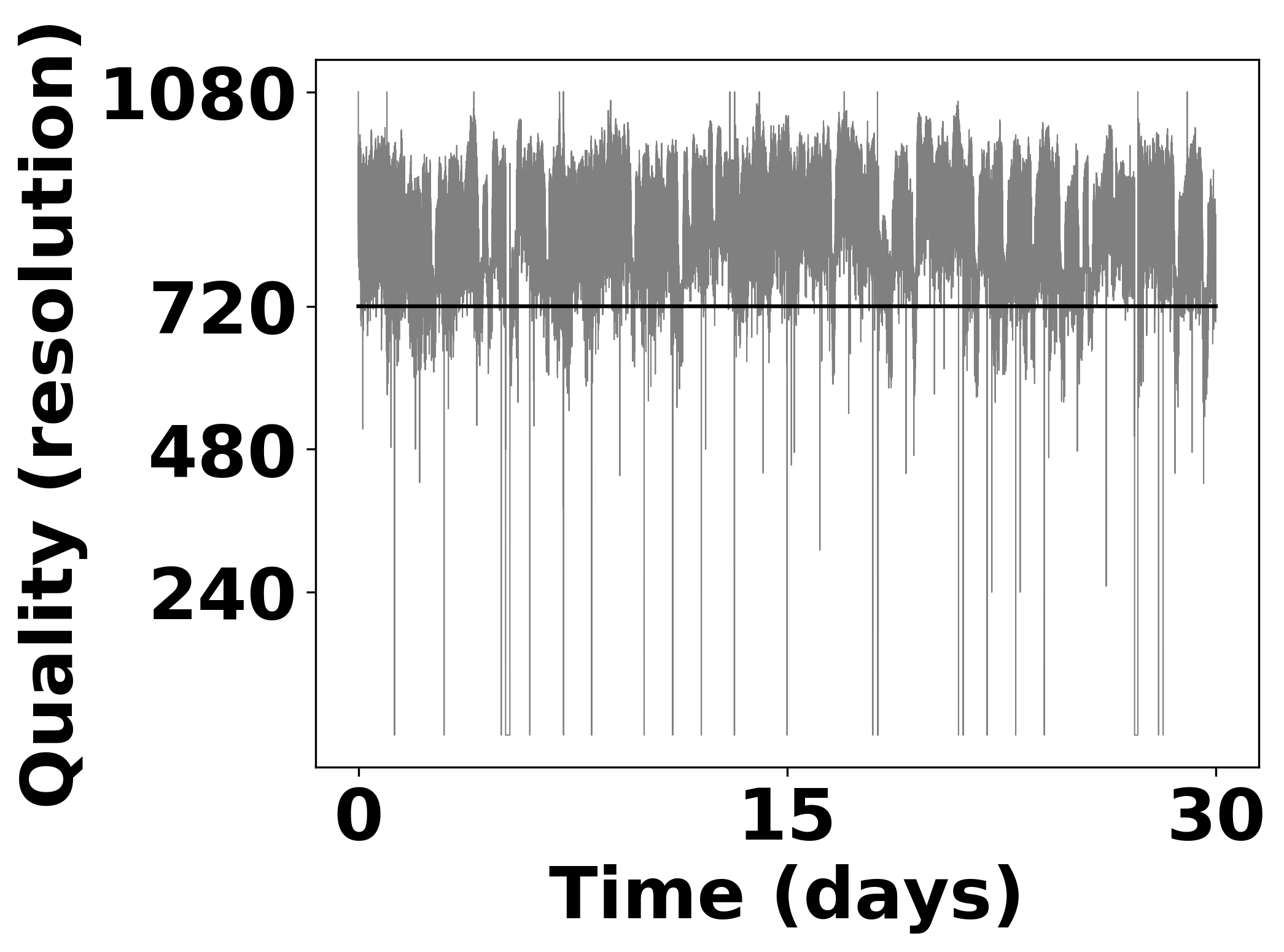}
  \caption{Client quality experience often falls below HD.}
  \label{fig:quality}
\end{figure}

UgoVor can also benefit both clients and content providers by enabling fine-grained analysis of the value clients receive for different quality tiers.
In current streaming systems, clients pay up front for service at a particular quality level.
For instance, Netflix offers service at three different tiers: basic quality for \$8.99, high-definition (HD) for \$12.99, or ultra HD for \$15.99.
Once clients have paid for service at some quality level, however, they have no guarantee that they will receive better service than a lower quality level.
Additionally, their service can be arbitrarily interrupted by buffering without consequence.
With UgoVor, however, these degradations in quality can easily be quantified and analyzed by both users and content providers.

Figure \ref{fig:quality} illustrates how clients in the Puffer data actually often receive lower quality than HD.
The flat line in the figure marks the minimum quality promised by an HD subscription to Netflix (for example).
The other line illustrates the actual quality experienced by a ``synthetic user'' calculated from the last whole month of the Puffer data.
This ``synthetic user'' captures the mean quality of every active stream at each point in time, thus representing the average quality experience of all clients at a given point in time.
As the figure shows, this quality varies widely and often falls well below HD.
This means that at any given time, many clients are often not receiving the service for which they have paid.

Using \tool, clients and content providers are able to monitor the content quality that users actually experience.
\tool further provides the ability to monitor rebuffering in real-time, enabling a new way of quantifying service quality.
On the client side, users could use this information to understand the true value of an HD subscription to streaming services as opposed to a standard (SD) subscription.
For some users, the actual quality they see with HD may be mostly the same as available with a cheaper SD subscription;
on the other hand, they may see higher quality on average, but with such frequent rebuffering as to outweigh the value of HD.
On the other side, service providers can use this fine-grained information to better understand patterns in their service.

To take this idea even further, \tool's precise monitoring enables a style of pricing where clients pay in real-time based on the quality of service they actually receive.
The basic idea is that the cost of streaming (e.g. a movie) can be split into chunks corresponding to the contract window, and clients pay for each chunk as they stream in real-time.
In this setting, the amount that a client pays can be informed by the satisfaction of the content's quality contract.
For instance, one pricing model might say that clients pay $N$ dollars for every chunk delivered at the expected quality, and any content provided with degraded quality is free (or discounted).
Rebuffering can also be accounted for in such models;
being as rebuffering events are explicit interruptions of service, it makes sense to account for these events in pricing as well.

%Hence, \tool's micro-auditing enables a new degree of service quality accountability that can greatly benefit clients and providers of streaming services.
%Service quality can be monitored in real-time not only in terms of content resolution, but also service interruptions due to rebuffering.
%With that information, clients can pay as they go for the service that they actually receive instead of merely the service that the content provider promises.
%Our analysis of the Puffer data indicates that because the service clients actually receive fluctuates significantly, this new level of accountability has the potential to offer clients substantial savings.

\section{Beyond Live Streaming and Other Extensions}
\label{extension}

\emph{Rewind/Forward.}
Our existing design and evaluation of \tool focus on live streaming. The
differentiating feature between live streaming services, such as streaming
TV, and streaming services such as Netflix is that the former, in general,
do not support rewinding or forwarding the video. With a few
modifications, \tool can support this necessary feature to accommodate the
popular streaming services. In particular, if the client's player rewinds
or forwards the video to some point that is not included in the client's
buffer, then this results in an ``out-of-order'' request to the server.
When such a request takes place, the client monitor resets its state and
notifies the auditor to do the same including starting a new contract
window. At the same time, the server monitor that sniffs the
``out-of-order'' request wipes off its buffer and also notifies the
auditor. In essence, moving the
playhead of the buffer outside the buffer results in restarting the
session from \tool's perspective no matter if the video moves backwards or
forward. At a first glance, the case where the requested playhead position
is in the client's buffer seems a bit more involved since it does not
result in extra requests that both monitors see. However, it is similarly
within small adaptations of the design of \tool and boils down 
to some extra communication between the parties to start a fresh
contract window. 
%See Appendix~\ref{appendixRF} for more details.
For rewinds, the main
challenge is that the \tool needs to be able to avoid detecting again
changes in the resolution or rebuffering events that it has already
reported. A simple fix is for the client monitor to ask the auditor to
restart its contract window. Hence, repeats of contract related events do
not affect contract checking.  In turn, the auditor can notify the server
monitor to adjust the start of the buffer to the chunk that is right
before the start of the current window so that all parties have a
consistent view.  For forward moves of the playhead inside the buffer,
again the same steps are sufficient; after the auditor and the server
monitor adjust their state they become again in sync with the client
monitor. The client has every incentive to comply and notify the auditor,
because otherwise it risks termination of the service due to
disagreements with the server monitor. 

\begin{figure}
    \centering
    \includegraphics[width=8cm, height=4cm]{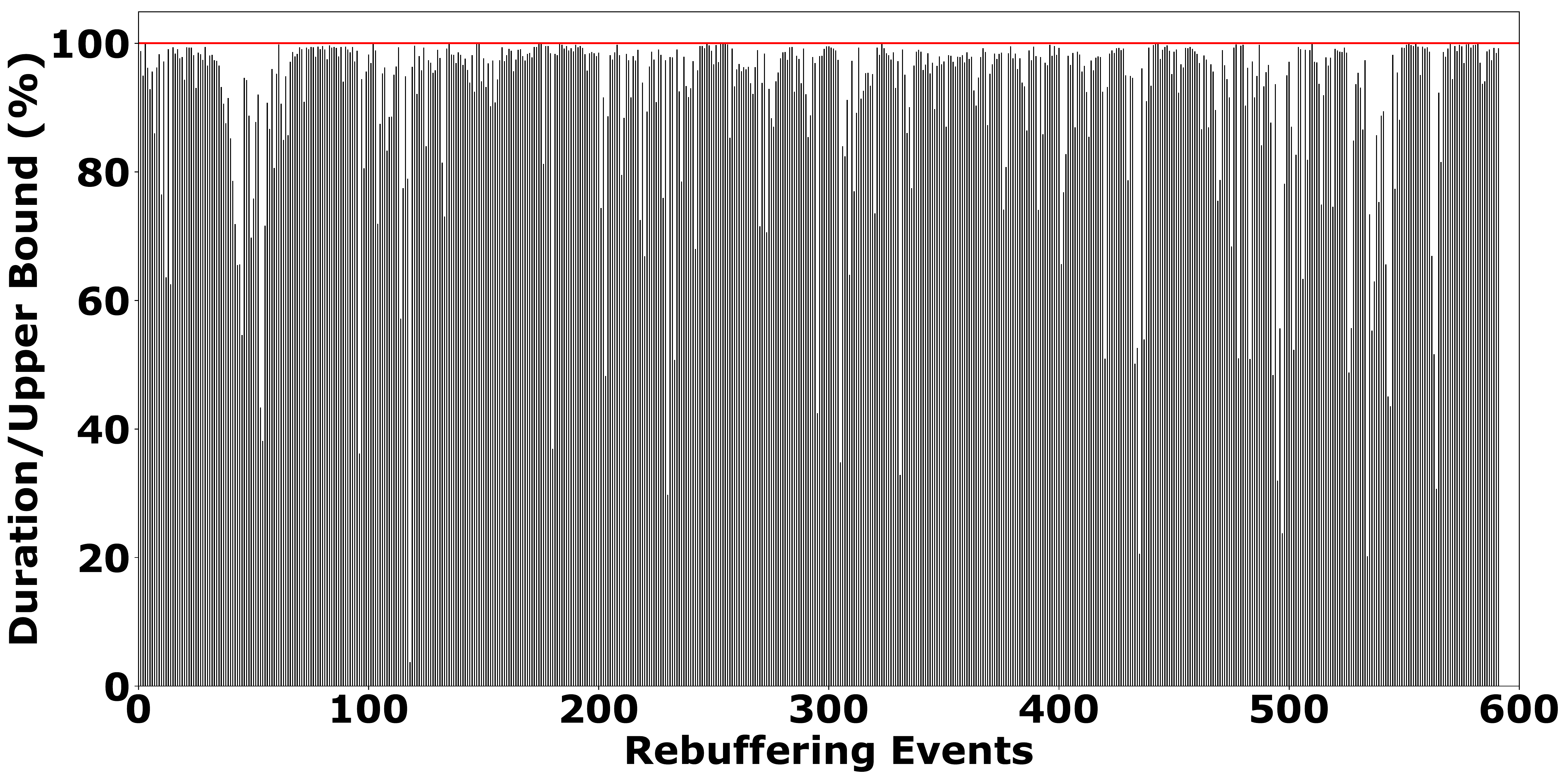}
  \caption{Ratio of rebuffering duration from the clients' perspective to the
  server's upper bound. 
%($c = 15 ms$). 
%AK removed because it implies the limit is artifically set to 15 ms.
The duration of all rebuffering
  events based on the clients is below the server's upper bound.}
  \label{fig:rebuf_bound}
\end{figure}

%\emph{Beyond Count of Rebuffering Events} 
\emph{Rebuffering Duration.}
\tool's contract language currently
only supports limiting the maximum count of rebuffering events and not 
their aggregate duration or their frequence or ratio over video chunk duration. 
Among these parameters, duration of rebuffering is the most challenging
one. However, \tool can safely monitor it with small modifications
to its contract language and design. Specifically, the server monitor can calculate an
upper bound for the duration of a rebuffering event it detects.
Concretely, if the 
rebuffering event occurs after the end of chunk $A$, then the upper bound is
$t^{ACK}_{B} - t_{A} - length(A) + c$ where $B$ is the next chunk the
server sends to the client, and $c$ is a constant
that corresponds to the delay for the client to place a new chunk it receives to 
the player's buffer. The client and server monitors can agree 
on the value of $c$ during \tool's bootstrap phase.
If the client monitor reports a
rebuffering event with duration $X$ that is below the upper bound the 
server monitor calculates, then the two parties are in agreement that the
 duration of the rebuffering event is $X$. Figure~\ref{fig:rebuf_bound} 
($c = 15 ms$) shows data that confirms the above theoretical exposition for
 our testbed.

 \emph{Monitoring Multiple Sessions.} \tool monitors a single streaming
 session at a time. However, all the events that the auditor 
 receives from the client and server monitors are verified. Thus auditors
 can collect data and make it available to tools that monitor
 contracts over multiple sessions, groups of clients, groups of servers,
 groups of CDNs etc.

\emph{Smart Clients.}
Our design assumes a client that downloads consecutive chunks of a video
file over a single connection.  Some players though can re-download chunks
at a higher resolution when there is spare
bandwidth~\cite{candid17nossdav}. \tool can accommodate such clients with
the server monitor synthesizing a worst case chunk from the re-downloaded
chucks (the lowest resolution, the earliest send timestamp and the latest
acknowledgment). The server monitor can also handle in the same manner
well-behaved clients that download multiple chunks in parallel. However,
with multiple connections between a client and a server, a dishonest
client can signal false contract violations to obtain for free
or low cost a few chunks from each connection. 
Nevertheless, the server monitor has the necessary information to detect such a client and 
notify auditors. Finally, same as for monitoring
multiple sessions, auditors can inform tools outside \tool 
about contract violations to detect dishonest smart clients that 
connect in parallel to multiple servers or CDNs. This is all
possible due \tool and the fine-grained  transparency
it enables.

\section{Discussion}
\label{discuss}

%CURRENT SYSTEM AND IT'S FEATURES AND ISSUES

\emph{A CDN cluster architecture with UgoVor.}
Deploying UgoVor in a CDN cluster could benefit from the existing typical
architecture of such clusters.  In particular, CDN clusters already
implement systems that perform live monitoring for single-ended
diagnostics and analytics.  Hence, piggybacking on this infrastructure, a
\tool server-side monitoring machine can serve a number of CDN servers in
a cluster, or across clusters. 
%Likewise, implementing sniffers on CDN servers requires only modular changes that do not affect their design. 

\emph{Partial deployment.}
\tool allows CDNs that deploy server-side monitors
to assess, via a virtual buffer, application-level client buffer health
metrics even when clients or auditors do not deploy \tool. This enables
CDNs to utilize fine-grained application-driven control, which is far more
effective than network-driven control, as we demonstrated in
Section~\ref{subsec:apps}.  Similarly, even in a scenario when not all the
clients deploy UgoVor, it remains feasible for content providers to
assess, with high statistical confidence, the aggregate quality
experienced by end-users in a region, via sampling (Appendix).

\emph{User Privacy.} The client monitor detects events of interest (such
as a rebuffering or a change in the resolution event) and notifies the
content-provider auditor about them.  Sharing this low-level information
is necessary, and strictly and significantly less than the amount of
information that content providers and, via them, third parties already
have, \emph{i.e.}, the semantics of the content users search
for and view. %Moreover, users in
%our system are incentivized to participate and share low-level performance information 
%because they get reimbursed when the service quality does not meet the agreed expectation.

{\bf Ethical Considerations.}
This work does not raise any ethical issues. The streaming
data set includes no personally-identifiable information, and no
user-identifiable or traceable information. For example, if the same user
utilized the Live TV streaming twice, this would show up as two
independent users in the trace. The trace does provide AS-level
information but we anonymized this information for the paper.

\section{Related Work}
\label{related}
\emph{Monitoring SLAs.} 
%{\bf Monitoring SLAs.}
Monitoring service level agreements and auditing networks has been an important topic for decades.
In \cite{Laskowski:2006}, the authors formally prove why accountability is essential for Internet sustainability.
Network tomography methods, \emph{e.g.}, \cite{Sommers:2007} utilize active or passive probing to 
measure metrics such as latency and packet loss rates in networks. 
\tool departs from such coarse-grained network auditing and focus on
end-to-end streaming \emph{micro}-monitoring.
%, which is relevant in terms of 
%agreements among CDNs, streaming infrastructures, content providers, \emph{etc}. 

\emph{Accountability.}
There has been been compelling work that addresses accountability in distributed systems~\cite{magpie-2003,X-Trace:2007,Pivot-Trace:2016,chen2017data,Haeberlen:2007} and networks~\cite{NSDI11:account}.
%In terms of accountability, 
%There has been work on 
In terms of loss and delay accountability, it was explored in the context of end-to-end flows~\cite{ArgyrakiMIAS07} and aggregate traffic passing between ASes~\cite{Argyraki:2010}. In particular, in~\cite{ArgyrakiMIAS07}, the authors propose an explicit accountability interface, through which ISPs report on their own performance, while in~\cite{Argyraki:2010} the authors propose a protocol that 
enables verifiable network-performance measurements. 
These proposals require audited systems to provide authenticated traffic
receipts to conclude which of the ASes is responsible for packet losses or
high latency and  have not been adopted or deployed. One of the reasons might be the high deployment threshold 
%(\emph{e.g.}, costly implementation 
%at AS border routers and clock synchronization), 
and lack of strong incentives. 
Clear economic incentives and a simplified, yet realistic, system model 
 is what sets \tool apart from prior work on accountability.

%\subsection{Service Auditing in Virtualized Environments}
\emph{Service Auditing in Virtualized Environments.} 
%{\bf Service Auditing in Virtualized Environments.}
A common problem in cloud computing is the inability for customers to understand the cost of their outsourced computation~\cite{Sekar:2011,Wang:2010,ParkHCP13}.
%that cloud-computing customers are facing is the inability to understand the cost of their outsourced computation~\cite{Sekar:2011,Wang:2010,ParkHCP13}.
The key issues in such systems are how to verify that a client actually utilized the resources it was charged for, and how to verify that the cost and the amount of available resources is in compliance with some agreed policy~\cite{Schnjakin:2010:CCA}. In analyzing and addressing the problems, researchers
considered 
the difficulty of fine-grained monitoring and the need for 
aggregation~\cite{Sekar:2011}, 
the need for third-party verifiers and witnesses~\cite{Wang:2010}, trusted hardware~\cite{ParkHCP13}, and attested reports~\cite{Sekar:2011}. %service emulation, \emph{etc}. 
Streaming contracts focus on an ecosystem that includes virtually (if not literally) all Internet end-users.  
%aim to address a much broader problem than accountability and micro-charging in a cloud-computing environment -- 
%our aim is to addresses a broader set of problems on the Internet and help heterogeneous network entities effectively contract among each other. 
Thus, the absence of strong user identities, which is a common case on today's Internet, is an unavoidable piece
of \tool. This, as well as the refrain from "heavy" mechanisms listed
above, distinguishes \tool from previously proposed solutions from
monitoring outsourced computation scenarios. 

\section{Conclusions}
\label{conclusions}

Video streaming is surging on the Internet, rapidly 
replacing traditional cable and satellite options.  To achieve high
performance, content providers outsource the content delivery to third
parties. Yet, content providers fundamentally lack direct insights into
the quality at which the clients are consuming their content.  In this
paper, we presented UgoVor, a system for micro-monitoring multilateral
streaming contracts between
clients, servers, and third-party auditors. The key to accurate
and scalable streaming contracts lies in removing the information
asymmetry present in existing streaming systems, by emulating a receiver's
video buffering status on the server side.  In addition, the key to
unleashing potential for incremental and wide deployment lies in strong
economic incentives and a simple tit-for-tat mechanism, which enforces
truthful endpoint reporting. 

By utilizing large-scale live streaming traces, we found the following.
($i$) Expectedly, streaming quality is far from ideal, prone to
low-quality resolution epochs and rebuffering events.  ($ii$) UgoVor can
capture resolution updates and re-buffering events, without incurring
inconsistent endpoint states, hence avoiding unnecessary service
disruption.  ($iii$) UgoVor is a scalable system, it does not affect the
streaming performance experienced by end-users and it adds a minimal
overhead.  ($iv$) UgoVor opens the doors to fine-grained streaming control
opportunities and it can help make performance-centric,
pay-what-you-experience, model a reality. Most importantly, it can help
shift the entire streaming ecosystem to a win-win-win spot: enabling the
much-sought transparency to content providers, giving an edge to
performant CDNs and brokers, and ultimately cutting the cost and improving
video-streaming performance for end users.

\bibliographystyle{plain}
\bibliography{reference,bib,pl}

\appendix
\section{Appendix: Ugovor Bootstrap}
\label{appendixBS}

%Our discussion about \tool so far assumes that 
%both the client and the server have deployed \tool and that their monitors
% ``know'' what is the contract between the client and the server.
%We also assume that they are both aware of the address of the auditor. 
%However, such assumptions are not realistic. Indeed, the common scenario
It is certainly possible that either the server or the client have not deployed \tool.
Furthermore, even if both server and client have deployed \tool, they need
a mechanism to agree on the contract and the auditor for their session. 

In terms of the contract and the auditor, our approach is that the server is the
one that specifies both and informs the client. After all, the server is the one 
that offers the service on behalf of the content provider, who is the
authority on the expected quality parameters of the service. Moreover, the 
users already make
offline agreements with a number of content providers when they subscribe
to their video service.  In other words,
the server's choice of the contract at the \tool level parallels the real life
legal contracts between the provider and the consumer of the service.
Simultaneously, at the 
technical level we avoid the off-line configuration of the client's
monitor with all the different
(and possibly changing) contracts different servers are supposed to live up to. 
For the same reason,
the server also selects the auditor and eliminates the need for 
offline  coordination between the client and the content provider.
Operationally, the server adds
two extra headers (details provided below) to the reply to the client's first HTTP(S) request for the
video. The client monitor picks up the contents of the two headers and
uses them to set up a connection with the auditor.

Furthermore, the initial request of the client 
also carries an extra header, proposing the use of UgoVor. If the server has
deployed \tool then the sniffer module recognizes the header and adds the
headers for ($i$) the contract and ($ii$) the IP of the auditor to
the server's reply to kick-off the \tool session. If the server has not deployed \tool then it ignores
the extra header of the request. In turn, the client monitor does not
receive the extra contract and auditor headers and thus it disengages
from the rest of the session.
Hence, monitoring only occurs if both client and server monitors are
in place; if either monitor is missing, \tool does nothing to interfere
with normal communication.
In this way, the design of \tool allows for the
interaction of clients and servers even if only some participants have deployed it. 
\section{Appendix: AS-Level Analysis}
\label{appendixAS}
Here, we aim to provide AS-level analysis. However, given that there are
more than 1,300 unique ASes associated with the streaming sessions in our
data set, we focus on the top 10 ASes in terms of the number of their
streaming sessions.  Each of
the ASes we selected has at least 9,000 streams. Importantly, the top
three ASes in terms of the number of streaming sessions are U.S. \emph{mobile
providers} in U.S., while the remaining ones are U.S. cable providers.
We explain below that the mobile vs. cable dimension has impact both on
users' behavior and streaming performance. In all scenarios, we use box
plots, where the box covers between 25\% and 75\% of the distribution.  In
addition, the "error bars" denote 5\% and 95\% of the distribution. Finally,
the black arrows indicate outliers, \emph{i.e.}, groups of
points that are far above the 95th percentile of the data.

\begin{figure*}[]
  \centering
  \begin{minipage}[t]{0.25\textwidth}
    \centering
    \captionsetup{justification=centering}
    \includegraphics[width=1.7in]{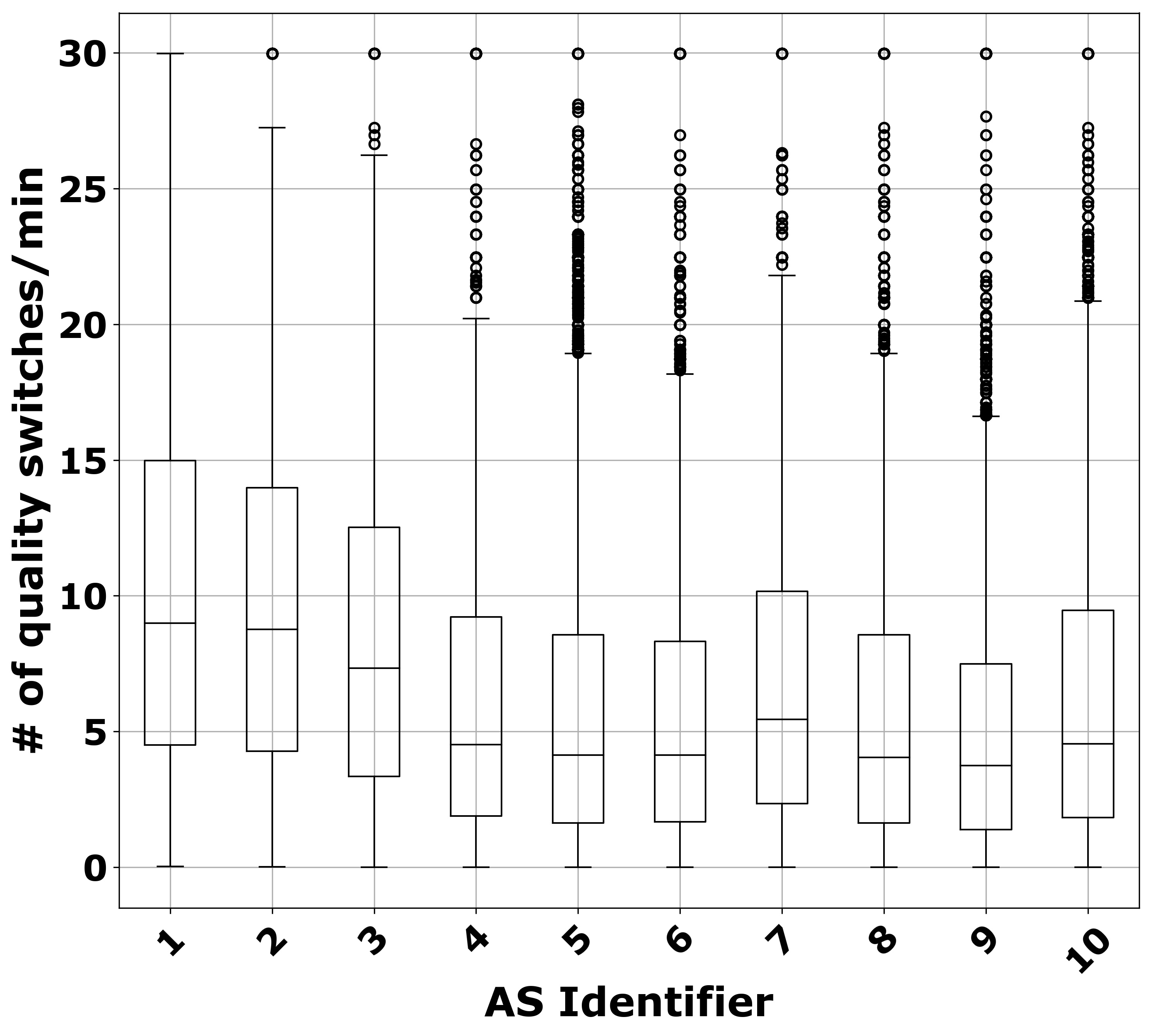}
  \end{minipage}\hfill
  \begin{minipage}[t]{0.25\textwidth}
    \centering
    \captionsetup{justification=centering}
    \includegraphics[width=1.7in]{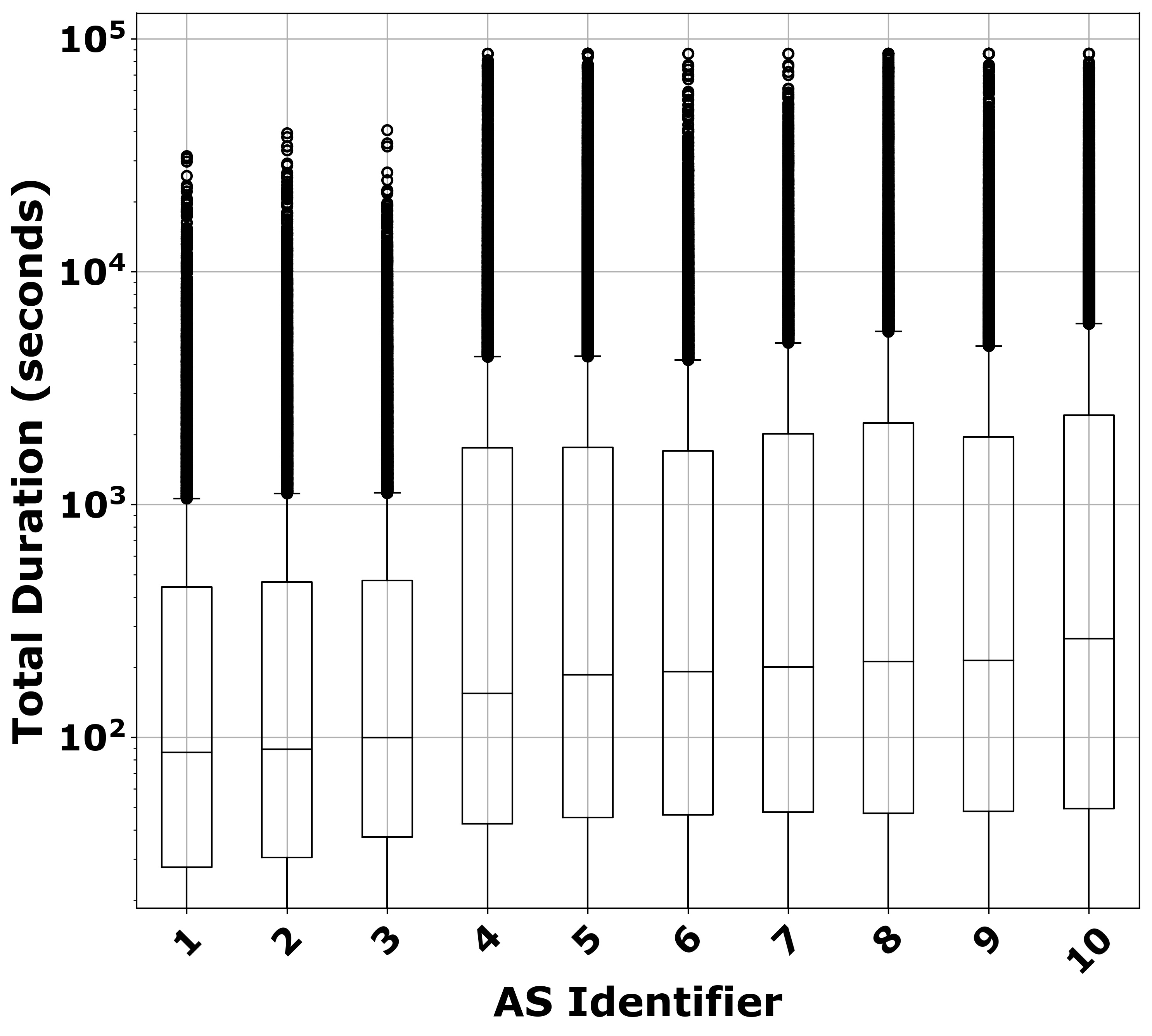}
  \end{minipage}\hfill
 \begin{minipage}[t]{0.25\textwidth}
    \centering
    \captionsetup{justification=centering}
    \includegraphics[width=1.7in]{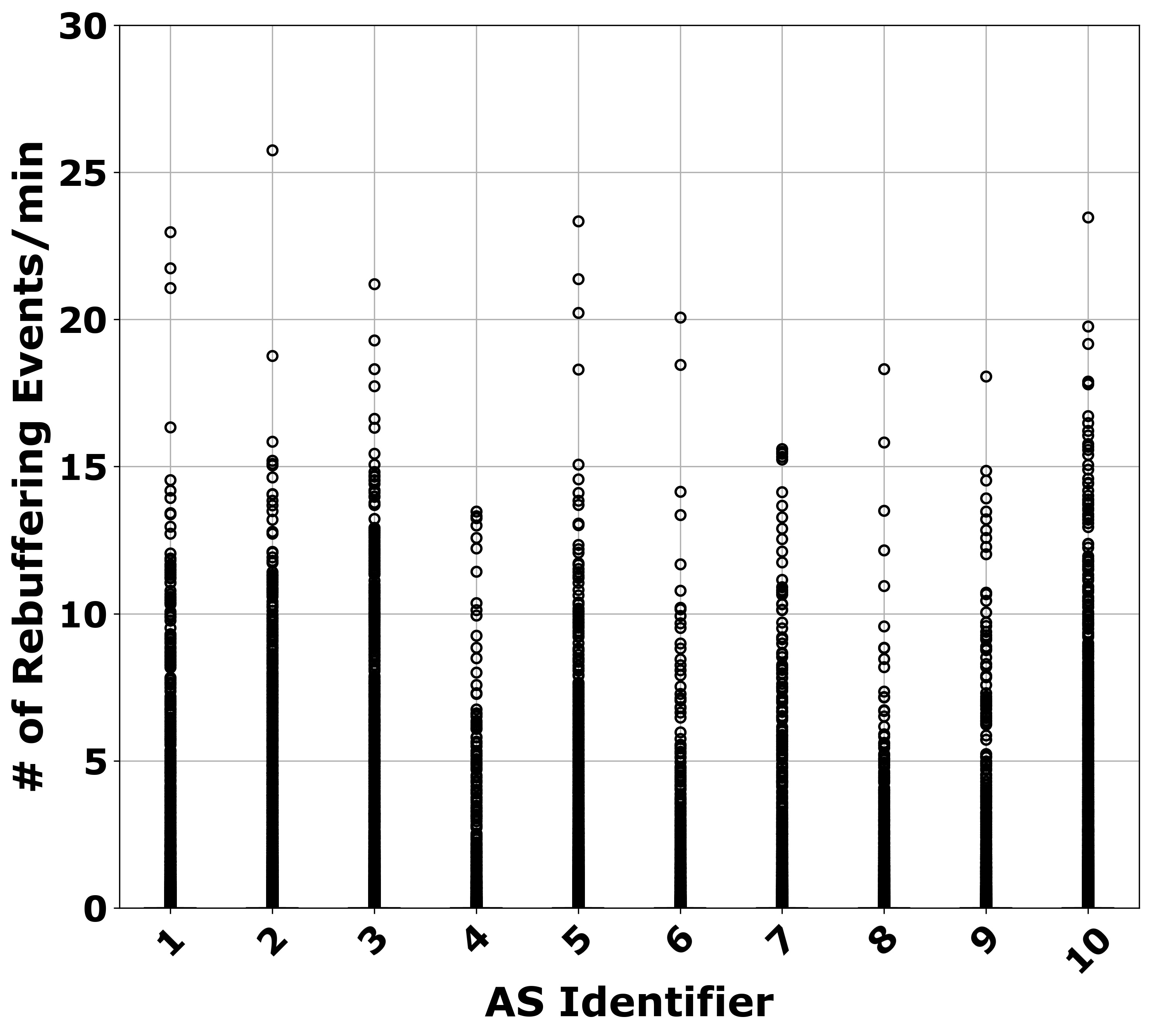}
  \end{minipage}\hfill 
 \begin{minipage}[t]{0.25\textwidth}
    \centering
    \captionsetup{justification=centering}
    \includegraphics[width=1.7in]{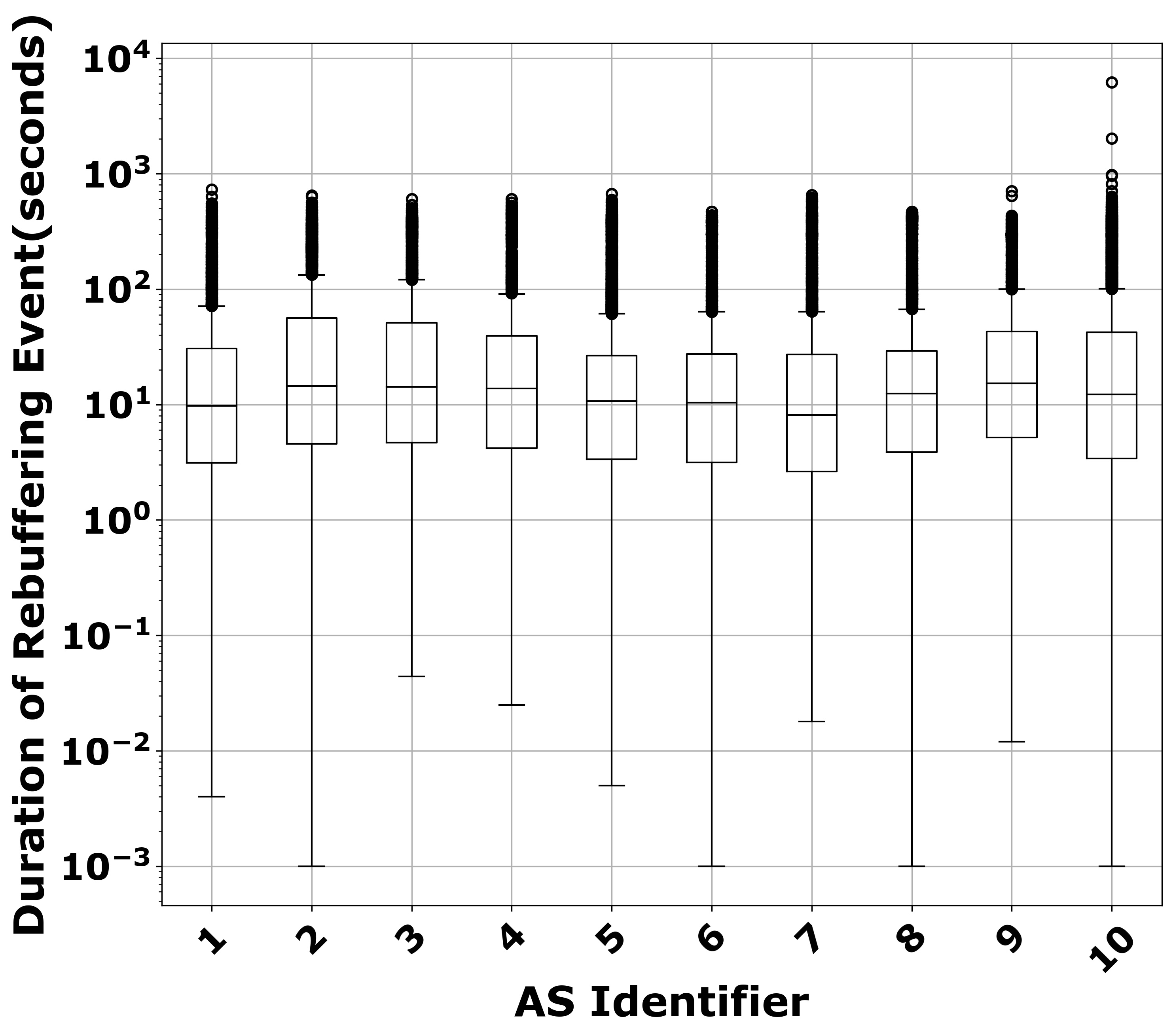}
  \end{minipage}
   \caption{Results over top 10 ASes,showing (from left to right): (a) Box plot of the number of video resolution switches per minute within a session, (b) box plot of the duration of streaming sessions, (c) box plot of the number of rebuffering events per minute, and (d) box plot of the duration of rebuffering events.} 
   \label{fig:ASes}
\end{figure*}

Figure~\ref{fig:ASes} shows the same results as those in
Figure~\ref{fig:aggregate}, only as box plots for the top 10 ASes.
Figure~\ref{fig:ASes}(a) and Figure~\ref{fig:ASes}(b) highlight the
difference induced by the type of the underlying access network,
\emph{i.e.}, mobile (ASes 1-3) vs. cable (ASes 4-10).
Figure~\ref{fig:ASes}(b) demonstrates that the length of a streaming session for
mobile ASes is approximately 5x shorter in the median case and approximately
10x shorter at 75\%. 
%Necessarily, 
We speculate that
given that the higher cost of using
mobile data, users are
possibly
more conservative when using it for streaming,
which is known to be a "data eater." While the cost is higher, the
streaming quality is not better though.

Figure~\ref{fig:ASes}(a) shows the number of streaming resolution changes
per minute for the ten ASes.  Again, we see a notable difference between
the top three mobile ASes, and the remaining seven. In particular, the
number of resolution changes per minute is approximately twice as high
in the median case for mobile providers. We
hypothesize that this comes
from the specific endpoint algorithm behavior. Puffer is an ML-scheme that
tries to learn the best streaming resolution to send data to the
client~\cite{puffer19}. Given the known latency and throughput variability
in mobile networks, the number of switches might be induced by it.
Finally, the results in Figure~\ref{fig:ASes}(c) and
Figure~\ref{fig:ASes}(d) depict that there are no significant AS-level
differences in terms of rebuffering events. 

\emph{Effects on UgoVor}: In terms of UgoVor, we see that there exists an
increased number of resolution changes for mobile networks. However, this
does not affect \tool's scalability, which is exactly the issue
we demonstrate next.

%\section{Appendix: Rewind/Forward the Playhead Position in the Client's Buffer}
%\label{appendixRF}
%\input{section/appendixRF.tex}
\section{Appendix: Statistical Analysis}
\label{appendix}

We have a relatively large population of streaming sessions, and we would like to select a statistically representative sample of that population.
Cochran's formula provides a simple way to calculate the necessary sample population size with the following assumptions:
\begin{itemize}
\item Uniform random sample selection
\item A desired confidence level: $1 - \alpha$
\item A desired margin of error: $\epsilon$
\end{itemize}

Then the number of samples to collect for a statistically representative sample set is:
\begin{equation}
n = Z_{\alpha/2}^2 \frac{p (1 - p)}{\epsilon^2}
\end{equation}
Where $p$ is a parameter of the distribution of the population. Since we do not know this value, we conservatively set it to 0.5 to maximize the required sample size.

Given that we want to have a confidence level of 95\% and margin of error of 5\%, we take:
\begin{itemize}
\item $\alpha = 0.05$
\item $\epsilon = 0.05$
\item $Z_{0.05/2} = Z_{0.025} = 1.96$ according to a standard Z table
\end{itemize}

\begin{equation}
\begin{split}
n &= 1.96^2 \frac{0.5^2}{0.05^2} \\
n &\approx 384
\end{split}
\end{equation}

Hence we select 384 samples.

%%%%%%%%%%%%%%%%%%%%%%%%%%%%%%%%%%%%%%%%%%%%%%%%%%%%%%%%%%%%%%%%%%%%%%%%%%%%%%%%
\end{document}